\documentclass[11pt, a4paper]{article} 

\usepackage[english]{babel} 

\usepackage{microtype} 

\usepackage{amsmath,amsfonts,amsthm} 

\usepackage[svgnames]{xcolor} 

\usepackage[small, labelfont=bf, up]{caption} 

\usepackage{booktabs} 

\usepackage{lastpage} 

\usepackage{graphicx} 

\usepackage{enumitem} 
\setlist{noitemsep} 

\usepackage{sectsty} 
\allsectionsfont{\usefont{OT1}{phv}{b}{n}} 

\usepackage{geometry} 

\usepackage[T1]{fontenc} 
\usepackage[utf8]{inputenc} 

\usepackage{XCharter} 

\usepackage{fancyhdr} 
\fancypagestyle{firstpage}{ 
	\fancyhf{}
}

\newcommand{\authorstyle}[1]{{\large\usefont{OT1}{phv}{b}{n}\color{DarkRed}#1}} 

\newcommand{\institution}[1]{{\scriptsize\usefont{OT1}{phv}{m}{sl}\color{Black}#1}} 

\usepackage{titling} 

\newcommand{\HorRule}{\color{DarkGoldenrod}\rule{\linewidth}{1pt}} 

\pretitle{
	\vspace{-30pt} 
	\noindent\HorRule\vspace{10pt} 
	\fontsize{20}{22}\usefont{OT1}{phv}{b}{n}\selectfont 
	\color{DarkRed} 
}

\posttitle{\par\vskip 15pt} 

\preauthor{} 

\postauthor{ 
	\vspace{10pt} 
	\par\noindent\HorRule 
	\vspace{20pt} 
}

\usepackage{lettrine} 
\usepackage{fix-cm}	

\usepackage{xstring} 

\usepackage[sorting=none]{biblatex}

\usepackage[autostyle=true]{csquotes} 

\usepackage[T1]{fontenc}
\usepackage[utf8]{inputenc}

\usepackage{graphics}
\usepackage[margin=0cm, font=small]{caption}
\usepackage{graphicx}
\usepackage{amsmath}
\usepackage{setspace}
\usepackage{relsize}
\usepackage[colorlinks = true, linkcolor = purple, urlcolor  = purple, citecolor = blue, anchorcolor = purple]{hyperref}
\usepackage{cleveref}

\usepackage{xcolor}
\usepackage{cuted}
\usepackage{xurl}
\usepackage{hyperref}
\PassOptionsToPackage{hyphens}{url}
\usepackage{siunitx}

\RequirePackage[utf8]{inputenc}
\RequirePackage{calc}
\RequirePackage{indentfirst}
\RequirePackage{fancyhdr}
\RequirePackage{graphicx,epstopdf}
\RequirePackage{lastpage}
\RequirePackage{ifthen}
\RequirePackage{lineno}
\RequirePackage{float}
\RequirePackage{amsmath}
\RequirePackage{setspace}
\RequirePackage{enumitem}
\RequirePackage{mathpazo}

\usepackage{multirow}

\usepackage{siunitx}
\usepackage{array}
\newcolumntype{C}[1]{>{\centering\arraybackslash}p{#1}}
\newcommand{\STAB}[1]{\begin{tabular}{@{}c@{}}#1\end{tabular}}

\definecolor{ao}{rgb}{0.0, 0.5, 0.0}

\title{
Beyond the Nucleus: Cytoplasmic Dominance in Follicular Thyroid Carcinoma Detection Using Single-Cell Raman Imaging Across Multiple Devices
}

\author{\scriptsize \noindent\authorstyle{Aurelien Pelissier$^{1,2,*}$, Kosuke Hashimoto$^{3,4,*}$, Kentaro Mochizuki$^{3,*}$, J. Nicholas Taylor$^{1,5}$, Jean-Emmanuel Clément$^{1}$, Yasuaki Kumamoto$^{3,6,7}$, Katsumasa Fujita$^{5,6,7}$, Yoshinori Harada$^{3,\dagger}$ and Tamiki Komatsuzaki$^{1,8,9,10,\dagger}$}\newline\newline
\textsuperscript{1}\institution{Research Center of Mathematics for Social Creativity, Research Institute for Electronic Science, Hokkaido University, Kita 20 Nishi 10, Kita-ku, Sapporo, 001--0020, Hokkaido, Japan}\\
\textsuperscript{2}\institution{Institute of Computational Life Sciences, Zürich University of Applied Sciences (ZHAW), 8820, Wädenswil, Switzerland}\\
\textsuperscript{3}\institution{Department of Pathology and Cell Regulation, Graduate School of Medical Science, Kyoto Prefectural University of Medicine, Kajii-cho, Kawaramachi-Hirokoji, Kamigyo, Kyoto, 602--8566, Kyoto, Japan}\\
\textsuperscript{4}\institution{Department of Biomedical Sciences, School of Biological and Environmental Sciences, Kwansei Gakuin University, 1 Gakuen, Uegahara, Sanda, Hyogo 669-1330 Japan}\\
\textsuperscript{5}\institution{Advanced Photonics and Biosensing Open Innovation Laboratory, AIST-Osaka University,Yamadaoka, Suita, 565--0871, Osaka, Japan}\\
\textsuperscript{6}\institution{Department of Applied Physics,Osaka University,2-1 Yamadaoka, Suita,565--0871,Osaka,Japan}\\
\textsuperscript{7}\institution{Institute for Open and Transdisciplinary Research Initiatives, Osaka University, Yamadaoka, Suita, 565--0871, Osaka, Japan}\\
\textsuperscript{8}\institution{Institute for Chemical Reaction Design and Discovery (WPI-ICReDD), Hokkaido University, Kita 21 Nishi 10, Kita-ku, Sapporo, 001--0021,Hokkaido, Japan}\\
\textsuperscript{9}\institution{Graduate School of Chemical Sciences and Engineering Materials Chemistry, and Engineering Course, Hokkaido University, Kita 13, Nishi 8, Kita-ku,Sapporo, 060--0812, Hokkaido, Japan}\\
\textsuperscript{10}\institution{The Institute of Scientific and Industrial Research,  Osaka University, Mihogaoka, Ibaraki, 8-1, Osaka, 567-0047, Japan}\\
\textsuperscript{*}\institution{K. Mochizuki, K. Hashimoto and A. Pelissier contributed equally to this work.}\\
\textsuperscript{$\dagger$}\institution{Corresponding authors: \href{tamiki@es.hokudai.ac.jp}{tamiki@es.hokudai.ac.jp}, \href{yoharada@koto.kpu-m.ac.jp}{yoharada@koto.kpu-m.ac.jp}}
}

\date{\vspace{-5ex}}

\begin{document}

\maketitle

\begin{center}
\begin{minipage}{14.6cm}
\vspace{-1cm}
\begin{center}
{\large \textbf{Abstract}}
\end{center}
\vspace{-0.2cm}

Cytological diagnosis of follicular thyroid carcinoma (FTC) is one of  major challenges in the field of endocrine oncology due to absence of evident morphological indicators. Morphological abnormalities in the nucleus are typically key indicators in cancer cytopathology and are attributed to a range of biochemical alterations in nuclear components. Consequently, Raman spectroscopy has been widely used to detect cancer in various cytological samples, often identifying biochemical changes prior to observable morphological alterations. However, in the case of FTC, cytoplasmic features such as carotenoids, cytochromes, and lipid droplets have shown greater diagnostic relevance compared to nuclear features. This study leverages single-cell Raman imaging to explore the spatial origin of diagnostic signals in FTC and normal thyroid (NT) cells, assessing the contributions of the nucleus and cytoplasm independently. Our results demonstrate that Raman spectra from the cytoplasmic region can distinguish between FTC and NT cells with an accuracy of 84\% under co-culture conditions, maintaining robustness across multiple devices. In contrast, classification based on nuclear spectra achieved only 53\% accuracy, suggesting that biochemical alterations in the cytoplasm play a more significant role in FTC detection than those in the nucleus. Our work elevates the promise of Raman-based cytopathology by providing complementary organelle-dependent information to traditional diagnostic methods and demonstrating transferability across different devices.

\end{minipage}

\end{center}

\vspace{0.6cm}







%

\section{Introduction}



\noindent Cytopathological diagnosis nowadays represents the gold-standard diagnostic methods for many form of cancers~\cite{mody2018diagnostic}. Key features in cancer cytopathology include nuclear morphological abnormalities, such as an increased nuclear-to-cytoplasmic ratio and/or irregular nuclear shape~\cite{zink2004nuclear, fischer2020nuclear, singh2022nuclear}. However, determining whether a tumor is benign or malignant based solely on these morphological features can be challenging, as the diagnosis may vary depending on the cytopathologist's interpretation.
To reduce the false positive or false negative rates in 
diagnostics,
there is a need to support the histocytological evaluations with information based on biochemical compositions. 
%
For example, molecular analyses such as immunohistochemistry \cite{duraiyan2012applications} and gene expression profiling \cite{khan2001classification} can reduce diagnostic variability.
%
%
Raman microscopy is one of the promising solutions to increase diagnostic reliability as it is a non-destructive technique capable of providing good molecular specificity and sensitivity, while requiring minimal sample preparation~\cite{ikeda2018raman}. In the last few decades, the number of Raman studies focused on oncology-based problems, and more generally on various tissue and cellular pathologies, has been growing progressively~\cite{santos2017raman,cui2018raman}. While Raman measurements for clinical applications are generally performed  with low magnification objectives to diagnose neoplastic tissues containing dozens to hundreds of cells~\cite{wang2022diagnosis, jabarkheel2022rapid}, the excitation wavelength in the visible range also allows a higher spatial resolution and can provide hyperspectral Raman images of individual cells at the sub-cellular level~\cite{palonpon2013raman,hamada2008raman}. The extraction of chemical and spatial information from sub-cellular components within individual cells has the potential to provide a more comprehensive understanding of underlying biological processes and improve the accuracy of clinical diagnoses~\cite{taylor2019high}.

A typical example that would benefit greatly from new clinical tools is the diagnosis of follicular lesions of the thyroid~\cite{sobrinho2011follicular}, which is known to be particularly difficult due to the lack of obvious morphological criteria for malignancy~\cite{wang2010detection}. The diagnosis of thyroid lesions has many variants~\cite{o2018raman}, each involving different risks and necessary treatments for patients: Follicular Thyroid Carcinoma (FTC), Papillary Thyroid Carcinoma (PTC), Anaplastic thyroid cancer (ATC), Medullary Thyroid Carcinoma (MTC) and Follicular Adenoma (that is not malignant). While cytopathologic diagnosis by fine needle aspiration biopsies is nowadays widely accepted as the initial step in the management of thyroid nodules~\cite{suen2002fine,dean2015fine}, a significant number of cases (roughly \SI{30}{\%}) are reported as follicular tumors of unknown malignant potential due to incomplete evidence of malignant features~\cite{cibas2009bethesda}. A growing part of the detected thyroid cancers are overdiagnosed and overtreated~\cite{jegerlehner2017overdiagnosis,leboulleux2016papillary}, which indicates that a significant proportion of unnecessary surgical solutions and further treatment such as thyroidectomy and radiotherapy could be avoided by supporting the traditional fine needle aspiration cytology diagnosis of thyroid lesions with biochemical composition information. In this context, several researchers have experimented the differentiation of thyroid follicular lesions with Raman microscopy to improve clinical diagnosis, utilizing both cell lines~\cite{taylor2019high,harris2009raman,lones2010discriminating,teixeira2009thyroid} and tissue sections~\cite{rau2017proof,rau2016raman,rau2019raman}. By employing feature extraction and dimensionality reduction methods, they have reported relatively high accuracies, ranging from 75\% to 95\% in distinguishing thyroid cancer. Interestingly, follicular thyroid carcinoma has been more effectively distinguished using Raman bands associated with cytoplasmic components, such as carotenoids, cytochromes, and lipid droplets~\cite{rau2017proof, taylor2019high, sbroscia2020thyroid, hayakawa2023lipid}. This contrasts with other types of thyroid cancer, which are typically characterized by an enrichment of DNA-rich components (e.g., O-P-O backbone stretching, nucleic acids, $\alpha$-helix)~\cite{harris2009raman}. Consequently, in the context of FTC, current observations do not align well with the general expectation that cancer cells contain elevated levels of nuclear material due to increased mitotic activity. The exact spatial origin of these signals, whether from the cytoplasm or the nucleus, thus remains unclear. Another crucial question is whether these Raman spectral features indicative of malignancy ---derived from an independent set containing solely either malignant or benign follicular cell lines with minimal differences in Raman signals--- can be transferred to a more clinically relevant environment where these cells are mixed and interacting.

Furthermore, these studies also highlighted significant variability in Raman spectra, particularly when measurements were taken on different days or under varying experimental conditions~\cite{rau2017proof}, raising concerns about the technique’s reliability for large-scale clinical applications. For clinical deployment, Raman measurements need to be reliable and repeatable (with the same sample) over a very wide range of experimental conditions. However, a major issue of Raman microscopy is its low signal intensity at each wavenumber, making it highly sensitive to minor fluctuations in experimental conditions, which often leads to inconsistent spectral results between different days or devices~\cite{guo2020comparability}. As a result, extensive calibration, preprocessing, and postprocessing steps are needed to mitigate these variations. Despite various approaches being proposed in the literature~\cite{butler2016using,pence2016clinical,guo2018extended}, a robust and reliable tool for Raman-based clinical diagnostics is still lacking~\cite{pence2016clinical}. Our team recently developed a batch correction technique for Raman hyperspectral images by subtracting background spectra of non-cellular components~\cite{taylor2023correction}, but it has yet to be validated in a diagnostic context.

In this article, we leverage single-cell Raman imaging of follicular thyroid carcinoma (FTC-133, RO82W-1 cell lines) and normal thyroid (NT) cells (Nthy-ori 3-1 cell line) to introduce three important contributions that bring Raman diagnosis closer to clinical application. Firstly, we highlight the effectiveness of our recently-developed extrinsic background correction (EBC) framework~\cite{taylor2023correction} in the context of FTC diagnosis, which minimizes experimental variations and enhances the robustness of Raman signals. We show that EBC largely improve the consistency of spectral intensities compared to conventional standardization methods, where the classification accuracy of FTC single cells across devices improved from 68\% to 81\% without and with the use of EBC, respectively. This approach addresses a primary obstacle to the clinical adoption of Raman microscopy, ensuring reliable results across different devices. Secondly, we demonstrate the high potential of Raman spectroscopy by isolating the cytoplasm from the nucleus in single cells, revealing the cytoplasm as the primary source of diagnostic information. By focusing on cytoplasmic information, we achieve higher accuracy compared to analyzing the single cell as the whole. This is in stark contrast to cytological approaches that predominantly focus on nuclear morphology in tissue samples.
Using feature selection approaches, we identify key wavenumbers associated with abnormal densities of lipids, cytochromes, and phenylalanine in FTC compared to the cytoplasm of NT cells. Interestingly, pixel-wise spectra analysis reveals that cytoplasms are heterogeneous environments with varying degrees of "malignancy (defined by Raman signals)" across their area. Finally, to validate our approach in a setting closer to clinical conditions, we establish a co-cultured system where FTC and NT cells interact. Our classifier, trained solely on independent FTC and NT cell lines, achieve 84\% accuracy in differentiating FTC from NT single cells in this co-cultured environment. This high accuracy, maintained under co-cultured conditions, underscores the practical applicability of our method. Our work not only provides complementary organelle-dependent information to traditional diagnostic methods but also demonstrates the transferability of Raman-based cytopathology across different devices, paving the way for more accurate and widely-applicable diagnostics in endocrine oncology.

\section{Results}

\subsection{Standardizing single cell Raman hyperspectral images with the extrinsic background correction method}

\begin{figure}[h!t]
    \centering
    \includegraphics[width=1\columnwidth]{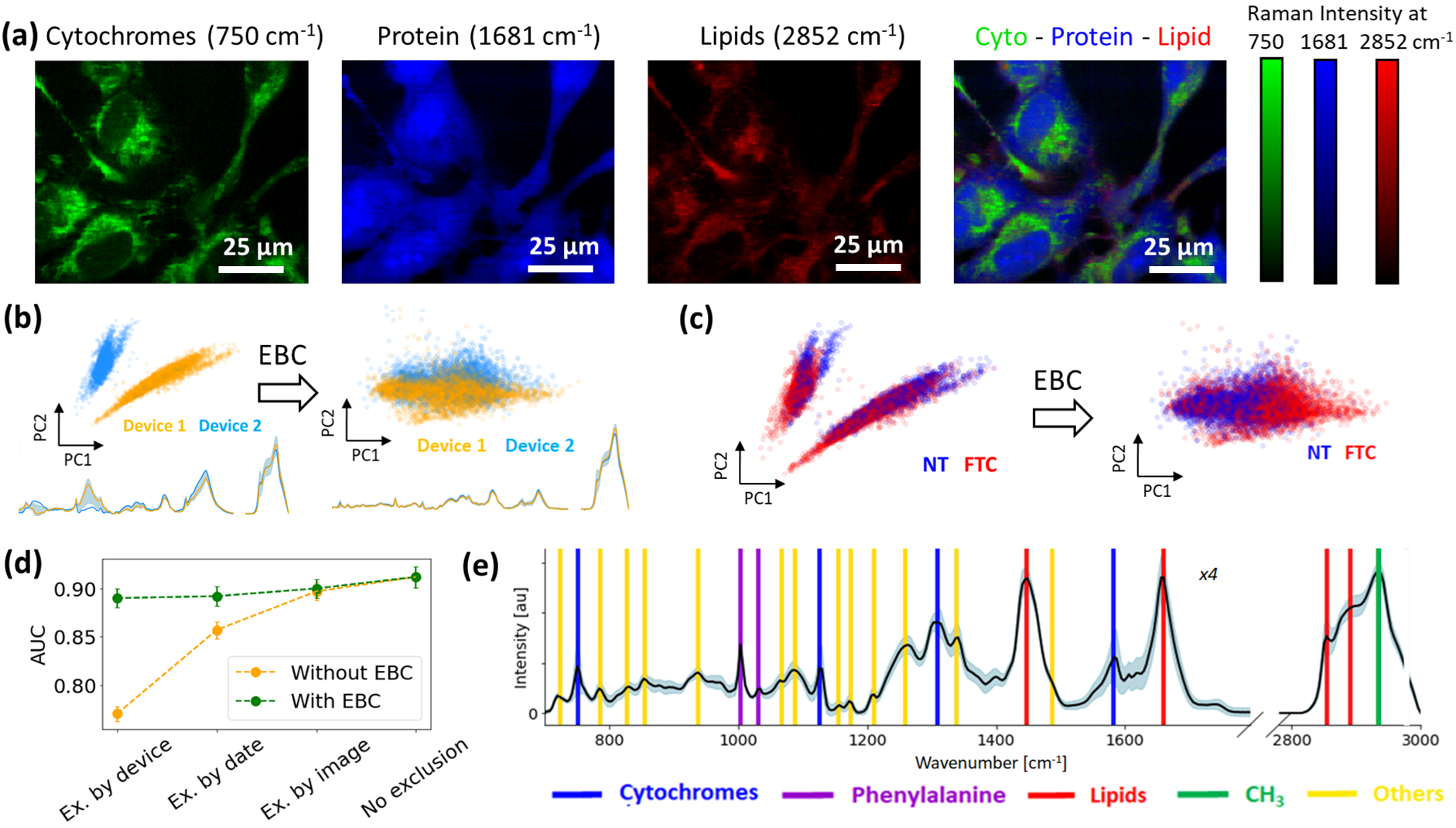}
    \caption{(a) A representative hyperspectral Raman image of mono-cultured FTC cells, illustrated with two-dimensional image frames at wavenumbers corresponding to cytochromes ($\SI{750}{cm^{-1}}$), proteins (amide I at $\SI{1681}{cm^{-1}}$), and lipids ($\text{CH}_2$ stretching at $\SI{2852}{cm^{-1}}$). (b \& c) Scatter plot of the first two principal components (capturing 88\% of the variance) of each spectrum labeled as (b) their device and (c) their malignancy with and without the extrinsic background correction (EBC) protocol~\cite{taylor2023correction}. The corresponding Raman spectra for averaged over each device is also shown. (d) Follicular thyroid carcinoma (FTC) vs  normal thyroid (NT) classification performance by excluding the cells measured with the same device, at the same date and from the same image, respectively. Classification was evaluated with a 10NN classifier on the first 5 principal components of the average cellular spectra with 5-fold cross validation averaged over 1000 sampling. The error bar represent one standard deviation. (e) Raman spectrum processed with EBC, averaged over all pixels within all images and devices. The gray area highlights the standard deviation. Major peaks in the spectrum are emphasized with vertical lines and colored according to their main chemical significances. The wavenumber region 700-$\SI{1800}{cm^{-1}}$ was increased by a factor of 4 to ease the presentation.  Broad band in 1410-$\SI{1470}{cm^{-1}}$ and 2800-$\SI{2900}{cm^{-1}}$ indicates the presence $\text{CH}_2$ bending and stretching, along with 1630-$\SI{1700}{cm^{-1}}$ for $\text{C=C}$ stretching, mostly contained in lipids~\cite{czamara2015raman}. 
    Narrow peak at $\SI{750}{cm^{-1}}$,$\SI{1127}{cm^{-1}}$,$\SI{1314}{cm^{-1}}$,$\SI{1585}{cm^{-1}}$ may be associated to cytochromes~\cite{okada2012label} in the mitochondria. Narrow peak at $\SI{1004}{cm^{-1}}$ and $\SI{1032}{cm^{-1}}$ may be matched with phenylalanine (C-C) stretching~\cite{rygula2013raman}. The \textit{silent region} (1800-$\SI{2800}{cm^{-1}})$ was cropped.}
    \label{fig:summary}
\end{figure}

\noindent Single-cell Raman images were obtained from FTC-133, RO82W-1 and SV40 T antigen immortalized~\cite{keeting1992development} Nthy-ori 3-1  cell lines. In short, measurements were performed on 16 different dates by two different devices (Supplementary Table \ref{table:setup}), resulting in 86 images. Hyperspectral Raman images data consist of three dimensional matrices, in which the first two axes corresponds to the image pixel positions and the third axis the wavenumbers along which the Raman spectrum is assigned at that pixel. One unprocessed image contains roughly \SI{1e5}{} spectra ($400 \times 250$ pixels) with $\sim$1000 values on the wavenumber axis, in the range of 700 - $\SI{3000}{cm^{-1}}$. Two-dimensional image frames at wavenumbers corresponding to cytochromes ($\SI{750}{cm^{-1}}$), proteins (Amide I, $\SI{1681}{cm^{-1}}$), and lipids ($\text{CH}_2$ stretching at $\SI{2852}{cm^{-1}}$), are shown for representative hyperspectral images in Figure~\ref{fig:summary}a. Raman spectra underwent standard preprocessing step, including cosmic rays removal, spatial-spectral denoising, area normalization and cellular segmentation (Methods Section~\ref{data_processing}). 
After segmenting single cells within each Raman image, we obtained 365 Nthy-ori 3-1 cells, 314 FTC-133 cells and 141 RO82W-1 cells (Total 820 cells).

As Raman microscopy is sensitive to devices optics and variability across cell cultures, we leveraged the extrinsic background correction (EBC)~\cite{taylor2023correction} to minimize inconsistency over different devices and experimental conditions (Methods Section~\ref{method:cell_identification}). As illustrated in Figure~\ref{fig:summary}b, spectra acquired from two different devices after conventional Raman preprocessing (denoising with singular value decomposition (SVD) and baseline correction with 7th order Polyfit) led to two very distinctive spectra distributions in the principal component (PC) space by visualizing the first two components. We also observed non-negligible discrepancies for spectra acquired on different dates of the same device (Supplementary Figure~\ref{fig:SI_dates}). On the other hand, spectra measured at different dates or from different devices overlapped well after incorporating EBC in the preprocessing. These observations were further supported by the classification performance of single cells into FTC and NT (Figure~\ref{fig:summary}c\&d, Supplementary Table~\ref{table:classif_prot}). Here, we applied a $k$-nearest neighbor ($k$NN) classifier on the average cell spectra to predict the given Raman signals as being FTC or NT, and found that EBC was beneficial when performing cell classification by excluding the measurements from the same device (accuracy improvement from 68\ to 81\%) as well as from the same date (accuracy improvement from 77\ to 81\%). We checked the dependency of the $k$ and $n$PCs for the $k$NN classifier and found for $1\le k \le 10$ and $1 \le n \le 10$ that unless $k<3$ or $n<3$ the classification performance was insensitive to the choice of the hyperparameters $k$ and $n$ (Supplementary Section~\ref{sup:kPC}, Figure~\ref{fig:k_nPC}). Hereinafter, we thus employ the 10NN classifier for the first five PCs space with 5-fold cross validation unless otherwise noted. These findings highlight the sensitivity of Raman microscopy to experimental conditions and the danger of mixing up Raman spectra across different devices or dates without appropriate preprocessing. Thus, in the following sections we only work with spectra that have undergone EBC processing.


We show in Figure~\ref{fig:summary}c the average Raman spectrum of acquired images obtained after processing with EBC. Several peaks, potentially holding significant biochemical implications for FTC diagnosis, can be discerned (Supplementary Table~\ref{table:wavenumbers}). 

\subsection{Identifying nucleus and cytoplasm spectra}

\begin{figure}[h!t]
    \centering
    \includegraphics[width=1\columnwidth]{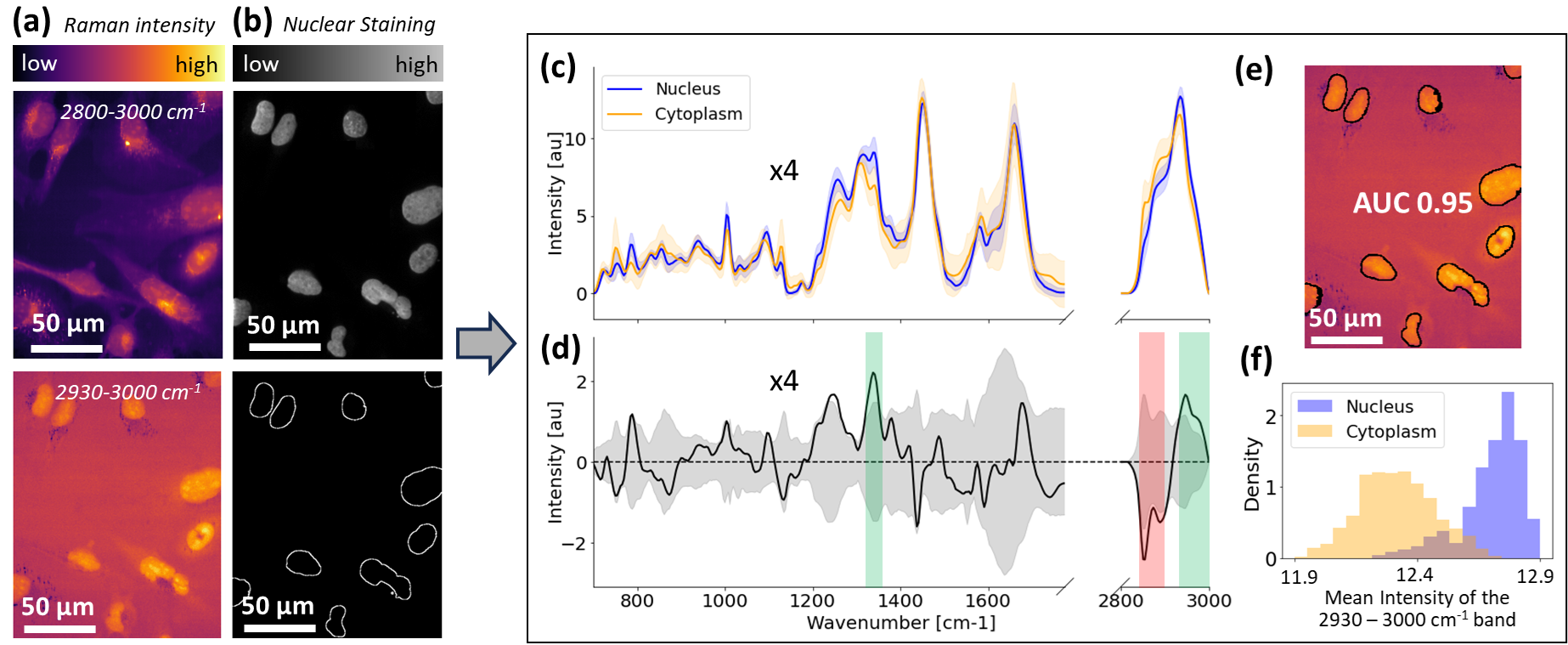}
    \caption{Nucleus-cytoplasm spectra identification in Single cell Raman images. (a) A representative Raman image (a co-joint culture of FTC and NT cells), illustrated with two-dimensional image frames of Raman intensity averaged over the $2800 - \SI{3000}{cm^{-1}}$ range (top), highlighting cell regions, and $2930 - \SI{3000}{cm^{-1}}$ range (bottom), highlighting potential nucleus regions. (b) Fluorescence images showing nuclear staining with TO-PRO-3 iodide, indicating precisely the nucleus regions in the representative image (top), along with the nucleus contours following basic image processing (bottom). (c) Averaged Raman spectra of nucleus and cytoplasm regions and (d) their average difference spectra, with shaded region representing one positive and negative standard deviation, where the Raman intensities are multiplied by a factor of 4 from $\SI{700}{cm^{-1}}$ to $\SI{1800}{cm^{-1}}$ for visual clarity. Wavenumber bands most relevant to the nucleus/cytoplasm differentiation are highlighted in green and red, where green indicate a positive difference and red a negative difference. (e) Raman image overlayed with nucleus contours from nuclear staining images, showing good alignment. Predicted nuclear spectra from the Raman image results in an AUC of 0.95 compared to the ground truth nucleus mask. (f) Distribution of average Raman intensity over the $2930 - \SI{3000}{cm^{-1}}$ band of cellular regions, showing a low overlap between nucleus and cytoplasm distributions.}
    \label{fig:nuc_identification}
\end{figure}

\noindent Single-cell Raman images offer detailed insights into subcellular structures, potentially uncovering relevant biochemical element localized in the cells~\cite{taylor2019high} (Figure~\ref{fig:nuc_identification}a). For example, analyzing Raman spectra from nucleus and cytoplasm regions distinctively was shown to yield valuable insights into the diagnosis of colon cancer by identifying information not discernible from standard whole-cell spectral analysis~\cite{liu2017raman}. 

To reliably discriminate nucleus from cytoplasm spectra in our images, we performed nuclear staining with TO-PRO-3 iodide (Figure~\ref{fig:nuc_identification}b) to obtain the nucleus localization of cells from 10 images containing FTC and NT cells (Methods~\ref{immaging}). We show Figures~\ref{fig:nuc_identification}c the resulting average spectra corresponding to both nucleus and cytoplasm region, and their difference on Figures~\ref{fig:nuc_identification}d. Here, the region of cytoplasm was defined by subtracting the nucleus region identified by the nuclear staining from the region of a single cell, which is the half of the clusters that exhibit the highest intensity in the high wavenumber region (2800-$\SI{3040}{cm^{-1}}$), as outlined in Method \ref{method:cell_identification}.  We quantified the wavenumber ability to discriminate nucleus from cytoplasm region with an ANOVA $F$-test, and found that the $1320 - \SI{1350}{cm^{-1}}$, $2840 - \SI{2910}{cm^{-1}}$ and $2930 - \SI{3000}{cm^{-1}}$ bands had the highest discriminating potential (Supplementary Figures~\ref{fig:ground_nuc_SI}a). Our findings are consistent with prior studies~\cite{carvalho2017raman, delfino2019multivariate}, which have identified these areas as effective discriminators between the nucleus and cytoplasm regions. Our analysis also revealed that the $2930 - \SI{3000}{cm^{-1}}$ range alone, corresponding to $\text{CH}_3$ stretching, was able to discriminate between nucleus and cytoplasm regions with an AUC of 0.95 (Figure~\ref{fig:nuc_identification}e,f). Using Raman intensities from other bands (addition, subtractions and ratios) did not improve the discrimination performances (Supplementary Figure~\ref{fig:ground_nuc_SI}b). Also, we did not find significant differences in nucleus/cytoplasm classification performances between NT and FTC cells. After optimizing our classifier parameters with these 10 ground truth images, we classified all spectra of our 820 cells into either cytoplasm and nucleus by $k$-means clustering on the average intensity of $\text{CH}_3$ stretching region 2930-$\SI{3000}{cm^{-1}}$, which was shown to be the best performance in classification among the above three bands (Methods~\ref{method:nucleus_identification}).

\subsection{Organelle dependency on differentiation between FTC and NT}
\begin{figure}[h!t]
    \centering
    \includegraphics[width=0.95\columnwidth]{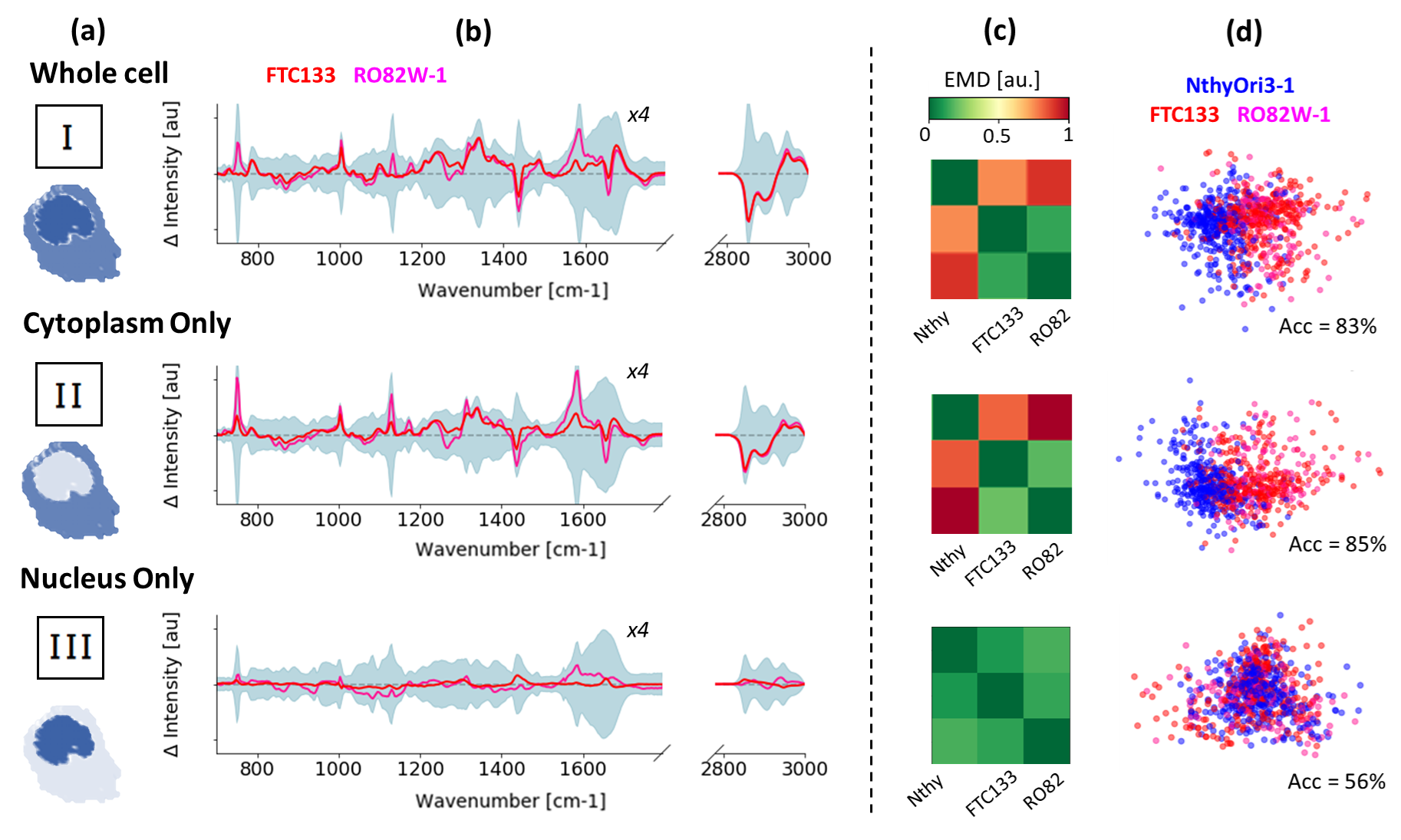}
    \caption{Organelle analysis of Nthy-ori 3-1, FTC-133 and RO82W-1 cell lines, involving the whole cell (I), cytoplasm region only (II), and nucleus region only (III). (a) Representative cell with its nucleus and cytoplasm pixels colored in dark blue and light blue, respectively. Pixels not considered in the analysis of nucleus and cytoplasms were shaded. (b) Averaged difference spectra of Raman spectra of Nthy-ori 3-1 minus those of FTC-133 and RO82W-1 with shaded region representing one positive and negative standard deviation, where the Raman intensities are multiplied by a factor of 4 from $\SI{700}{cm^{-1}}$ to $\SI{1800}{cm^{-1}}$ for visual clarity. (c) Earth Mover’s Distance (EMD) distance matrix between the three spectra distributions, normalized by its maximum value.. (d) Pixel-wise PC1-PC2 (contains 88\% of the information) scatter plot, with each point corresponding the spectra of a single cell, averaged over the region of interest (whole cell, cytoplasm, nucleus), colored according to their malignancy (blue for Nthy-ori 3-1, red for FTC133 and pink for RO82W-1). The 5-fold cross-validation classification accuracy (Acc) with 10 nearest neighbors is also given (averaged over 1000 sampling).}
    \label{fig:region}
\end{figure}

\noindent Figures \ref{fig:region}a(I) depicts the spectral difference between Nthy-ori 3-1 and FTC cell lines (FTC-133 and RO82W-1) averaged over the whole cellular region.  Although discernible variations in spectral attributes are observed at particular wavenumbers, it remains uncertain whether the Raman spectroscopic data pertinent to FTC diagnosis is localized within specific subcellular compartments or, conversely, uniformly distributed throughout the cell. To elucidate this, we investigated the spectral difference between Nthy-ori 3-1 and FTC cell lines (FTC-133 and RO82W-1) averaged over nucleus region only (Figures~\ref{fig:region}IIb), and cytoplasm region only (Figures~\ref{fig:region}IIIb), whose regions are identified based on the discussion in the Section 2.2. 


As a metric to quantify how two sets of spectra differ from each other, we computed the Earth Mover's Distance (EMD)~\cite{rubner1998metric} to quantify distances between two discrete distributions. In short, the more two groups of Raman spectra resemble to each other, the smaller the EMD is. We note that EMD is computed directly in the wavenumbers space (whole spectra) and not in the PC space. 
Figures \ref{fig:region}c depicts the scatter plots in the projection of PC1-PC2 space (capturing 88\% of the variance) of pixel-wise Raman spectra, among with the EMD matrices between the three cell lines distributions in the wavenumber space.

As observed in the scatter plot, the distinction between the FTC and NT distributions is very poor for nucleus regions (EMD distance around 0.15 (a.u.), as indicated by the green heatmap in Figure~\ref{fig:region}IIIc) compared to those of cytoplasm ($>0.6$ a.u., i.e., the orange and red heatmap in Figure~\ref{fig:region}Ic) and whole cell region ($>0.6$ a.u., i.e., the orange and red heatmap in Figure~\ref{fig:region}IIc), which indicates that NT and FTC nucleus may contain less relevant chemical differences relative to their cytoplasm for differentiation between FTC and NT. As expected, the spectral differences at wavenumbers $\SI{750}{cm^{-1}}$, $\SI{1127}{cm^{-1}}$, $\SI{1314}{cm^{-1}}$ and $\SI{1585}{cm^{-1}}$ are amplified for the region assigned as cytoplasm relative to the whole cell region (c.f., Figure~\ref{fig:region}Ib and IIb), because these peaks can be mainly assigned to mitochondrial cytochromes~\cite{lennarz2013encyclopedia}. As for the differences between the FTC-133 and RO82W-1 cell lines, their spectral differences exhibit a similar trend, although the differences from RO82W-1 to Nthy-ori 3-1 are more pronounced than those from FTC-133. However, their spectral contribution in the high-wavenumber region is very similar, and thus, their corresponding EMD distance is still relatively small (See off-diagonal elements between FTC-133 and RO82W-1 in  Figure~\ref{fig:region}c).


As for single cell classification performances, we computed, for each single cell, three different averaged spectra within the region of interest (whole cell, cytoplasm, nucleus), for which the projected spectra in PC1-PC2 space are depicted in Figure\ref{fig:region}d. We evaluated the classification performance with a 10NN classification in the first five PC space.  

Here, the classification performance was quantified in terms of accuracy, 
AUC \cite{powers2011evaluation}, and 
F1score \cite{powers2011evaluation} (the latter two measures take into account uneven class distribution). As shown in Table~\ref{table:classif}
the FTC discrimination accuracy is similar using the Raman spectra averaged over the for whole cells (e.g., AUC 0.90) and those over the cytoplasm only regions (AUC 0.93), but very poor using those over the nucleus region (AUC 0.60). Values for F1 scores and accuracy show a similar trend.
This is in a good agreement with the difference spectra and pixel-wise PC1-PC2 plot in Figure~\ref{fig:region}b and Figure~\ref{fig:region}c. 


\def\arraystretch{1.2}
\begin{table*}[ht]
\centering
\caption{Cellular classification accuracy, AUC and F1 score with a 10 nearest neighbor classifier on the first 5 principal components of the average cellular spectra, computed by considering the whole cell, cytoplasm region only and nucleus region only. Interquartile range (IQR) are calculated by modifying the train/test split 1000 times and computed as half the difference between the third and first quartiles.}
\begin{tabular}{|C{4cm}|C{3cm}|C{3cm}|C{3cm}|}  
    \cline{2-4}
    \multicolumn{1}{c|}{} & Full cell & Cytoplasm only & Nucleus Only\\  
    \hline
    Accuracy & 83 $\pm$ 1.2\% & 85 $\pm$ 1.2\% & 56 $\pm$ 1.0\%\\ 
    \hline
    F1 score & 0.84 $\pm$ 0.011 & 0.86 $\pm$ 0.011 & 0.58 $\pm$ 0.008 \\
    \hline
    AUC & 0.91 $\pm$ 0.008 & 0.94 $\pm$ 0.008 & 0.60 $\pm$ 0.006 \\
    \hline
\end{tabular}
\label{table:classif}
\end{table*}

\def\arraystretch{1.}

\begin{figure}[h!t]
    \centering
    \includegraphics[width=1.0\columnwidth]{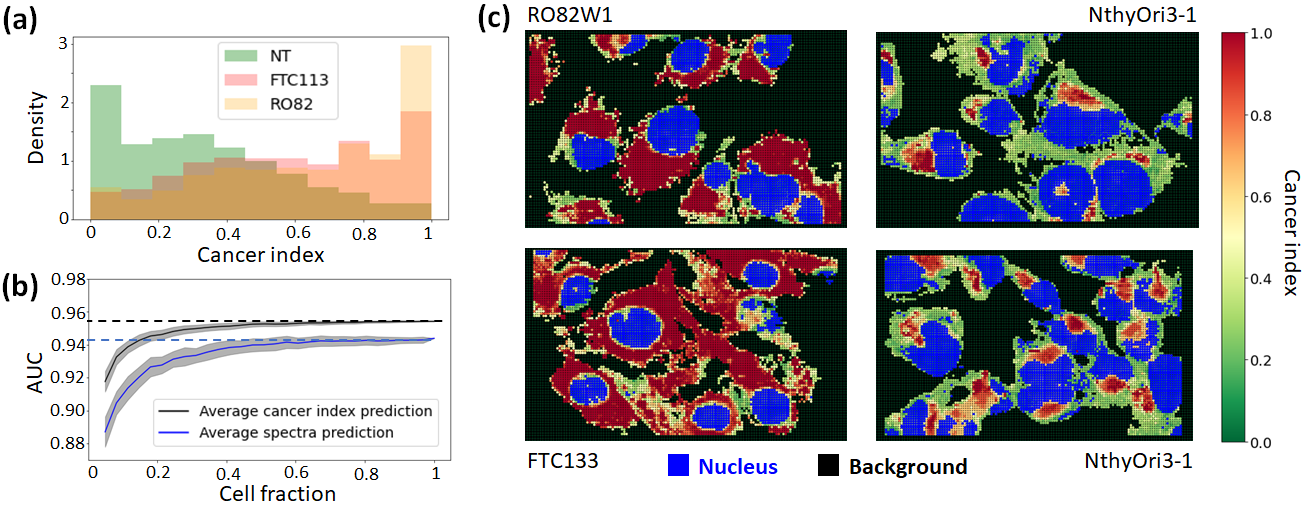}
    \caption{(a) Cancer index density distribution of all pixels in the cytoplasm of each cell line. (b) Classification performances (AUC) with partial cell information for prediction made with the average cytoplasm spectra (blue) and average cancer index (black). The AUC is averaged over 1000 resampling and the shaded area represents one positive and negative standard deviation. (c) Spatial distribution of the cancer index at each pixel on representative images of RO82W1, FTC-133 and Nthy-ori 3-1 cell lines. Pixels from non cell regions (background) and nucleus regions are colored in black and blue, respectively.}
    \label{fig:heterogenous}
\end{figure}

Next, to further analyze the subcellular compartments of FTC and NT cells,  we associated a cancer index to each pixel in each cell's cytoplasm, representing the ''cancer degree'' of that particular pixel. The cancer index is defined as the fraction of cancer neighbors, evaluated with a 10-nearest neighbor classifier in the first five PCs space of cytoplasm Raman spectra: the closer the index to 1 (0), the higher the cytoplasmic chemical micro-environment is similar to those of FTC (NT).  The distribution of cancer index values across pixels in cytoplasmic space for the representative images of FTC (FTC-133, RO82W1) and NT (Nthy-ori 3-1) systems are shown in Figure~\ref{fig:heterogenous}. Note that we focus on the cytoplasm only, as was already shown that nucleus carries less information for discrimination of FTC and NT.

As shown in Figure~\ref{fig:heterogenous}a, the cancer index distribution of FTC and NT cytoplasms differ significantly, with NT cell being heavily skewed toward low cancer index (mean cancer index for NT cytoplasms = 0.28) and vice versa (mean cancer index for FTC cytoplasms = 0.64).
As cells are heterogeneous environments containing various organelles,
different part of the cytoplasms may have different cancer index values.
As an example, pixels belonging to the outer part of the cytoplasm (membrane), were of low cancer index in FTC cells (Figure~\ref{fig:heterogenous}c). Lipids droplets have been observed in the nucleoplasm of some cells~\cite{farese2016lipid,ohsaki2016pml}, and were associated to high cancer index area in NT cells. 


As an alternative approach for the diagnosis of FTC cells, we predicted the cell's malignancy as the mean cancer index over its cytoplasm pixels, rather than with its average spectra as performed in the previous section. Performances were found to be slightly increased (AUC improvement from 0.94 $\pm$ 0.008 to 0.96 $\pm$ 0.007). Interestingly, the malignancy of the cells were predictable without measuring their entire cytoplasm area, with a nearly equivalent AUC (\SI{1}{\%} decrease) obtained by considering just one quarter of randomly sampled pixels in the cytoplasm area on average (Figure~\ref{fig:heterogenous}b). 
This indicates that the cell's discrimination could be performed without requiring the acquisition of the entire image, with a measurement of 25-50\% of the cell’s cytoplasm leading to a similar classification quality than the whole cytoplasm. 



\subsection{Transferability of information acquired on independent FTC and NT system to FTC-NT co-cultured system}

\begin{figure}[h!t]
    \centering
    \includegraphics[width=1\columnwidth]{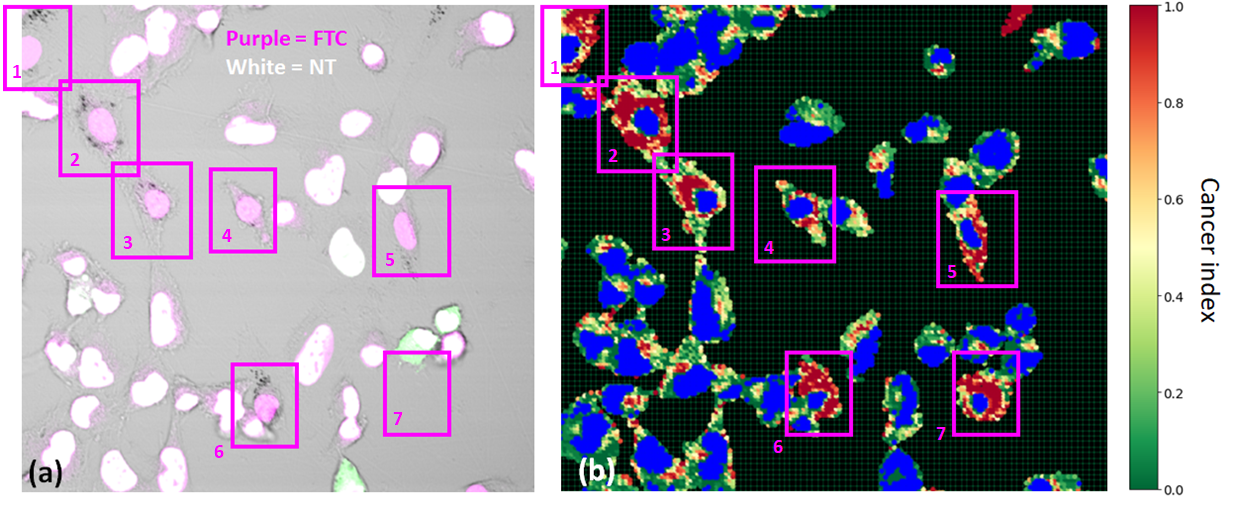}
    \caption{A representative image of the co-culture system of FTC and NT. (a) An overlaid microscopic fluorescence image of SV40 large T protein (green) and double strand DNA (magenta). Since Nthy-ori 3-1 should be positive for both, the color of nuclei in Nthy-ori 3-1 is white as a result of the overlay, in contrast, FTC cell is negative for SV40, which makes nuclei of FTC cells colored in magenta. Cells in which green color around white indicates dividing Nthy-ori 3-1. (b) A distribution of the cancer index at each pixel trained by only cytoplasm Raman signals, pixels from non cell regions and nucleus regions are colored in black and blue, respectively. The cancer cells are framed by a purple rectangle, note that the cell number 7 is missing because it was washed out during the cleaning process prior to the fluorescence measurement.}
    \label{fig:cancer_index}
\end{figure}

\noindent The immune system has the capability to detect and eliminate certain tumors before they progress to malignancy~\cite{pandya2016immune}. In clinical applications, it is unrealistic to assume that all cells in a given sample would be of the same class, i.e., cancer or noncancer. It is thus desirable to establish a mixed co-culture system that contains both classes, to confirm that our classifier constructed from independent and isolated cell lines is transferable to such a co-cultured system where each cell interacts with each other. We developed a mixed co-cultured system composed of FTC (FTC-133 or RO82W-1) and NT (Nthy-ori 3-1) cell lines. After the Raman measurements, FTC cells from NT cells were discriminated by independent fluorescence measurement (Figure~\ref{fig:cancer_index}a) of chemically injected SV40 large T antigen protein. Control experiments demonstrated that immunofluorescence imaging allow for the distinction of FTC/NT cells with 99.86\% accuracy (Supplementary Section~\ref{SV40}).

We acquired 6 hyperspectral Raman images of Nthy-ori 3-1/FTC-133 and 3 of Nthy-ori 3-1/RO82W-1 mixed co-cultured systems. After segmentation of single cells, we were left with 480 cells across the 9 images. To predict the malignancy of each cell, we calculated the average cancer index across their cytoplasmic areas, applying a threshold of 0.58. This threshold was selected to maximize the F1 score of our training data. Then, defining ground truth from position matching to SV40-staining fluorescence images (Figure~\ref{fig:cancer_index}, Supplementary Figure~\ref{fig:co-culture1}), we found that 84\% of the cells were correctly classified on average in the co-cultured data (Table~\ref{table:co-culture}), which is in good agreement with the 85\% found in our training data of the mono-cultured systems, either FTC-133, ROW2W-1, or Nthy-ori 3-1. Note that, among the 9 images, one particular image (RO82\_Nthy\_no2 in  Figure~\ref{fig:co-culture1}) yielded a poor classification whose accuracy is 48\%, in which the number of false positively predicted cells --- i.e., even though cells are actually noncancerous, the classifier predicts as cancerous --- is 18 out of 36 Nthy-ori 3-1 cells. We confirmed that this apparent outlier was robust to our choice of classifier, subset of wavenumbers in Raman spectra, and other variants of background correction schemes (Note here that the falsely negative predicted cells, i.e., the number of cells the classifier predicts as non-cancerous even though they are actually cancerous, is 3 in RO82\_Nthy\_no2, which is a similar rate as the other images). A possible interpretation is that cells in co-cultured environments may interact, leading to alterations in the state of non-cancerous cells due to their proximity to cancerous cells \cite{ivers2014dynamic}, such as through mitochondrial dysfunction and Ras signaling \cite{ohsawa2012mitochondrial}. It should be noted that, for the aim of detecting the existence of cancerous cells, such false positive prediction of non-cancerous cells does not cause problem in practice because detection of non-cancerous cells that interact with cancerous cells means the same as detection of cancerous cells.

As demonstrated in the previous section, nucleus carries less information for the differentiation of FTC and NT diagnosis, and the same analysis using nucleus region yielded a poor cell diagnosis accuracy with only 53\% accuracy (Supplementary Figure~\ref{fig:cancer_index_nuc}). 

\newpage

\subsection{An optimized set of Raman shifts for differentiation between FTC and NT's cytoplasm}

\begin{figure}[h!t]
    \centering
    \includegraphics[width=1.0\columnwidth]{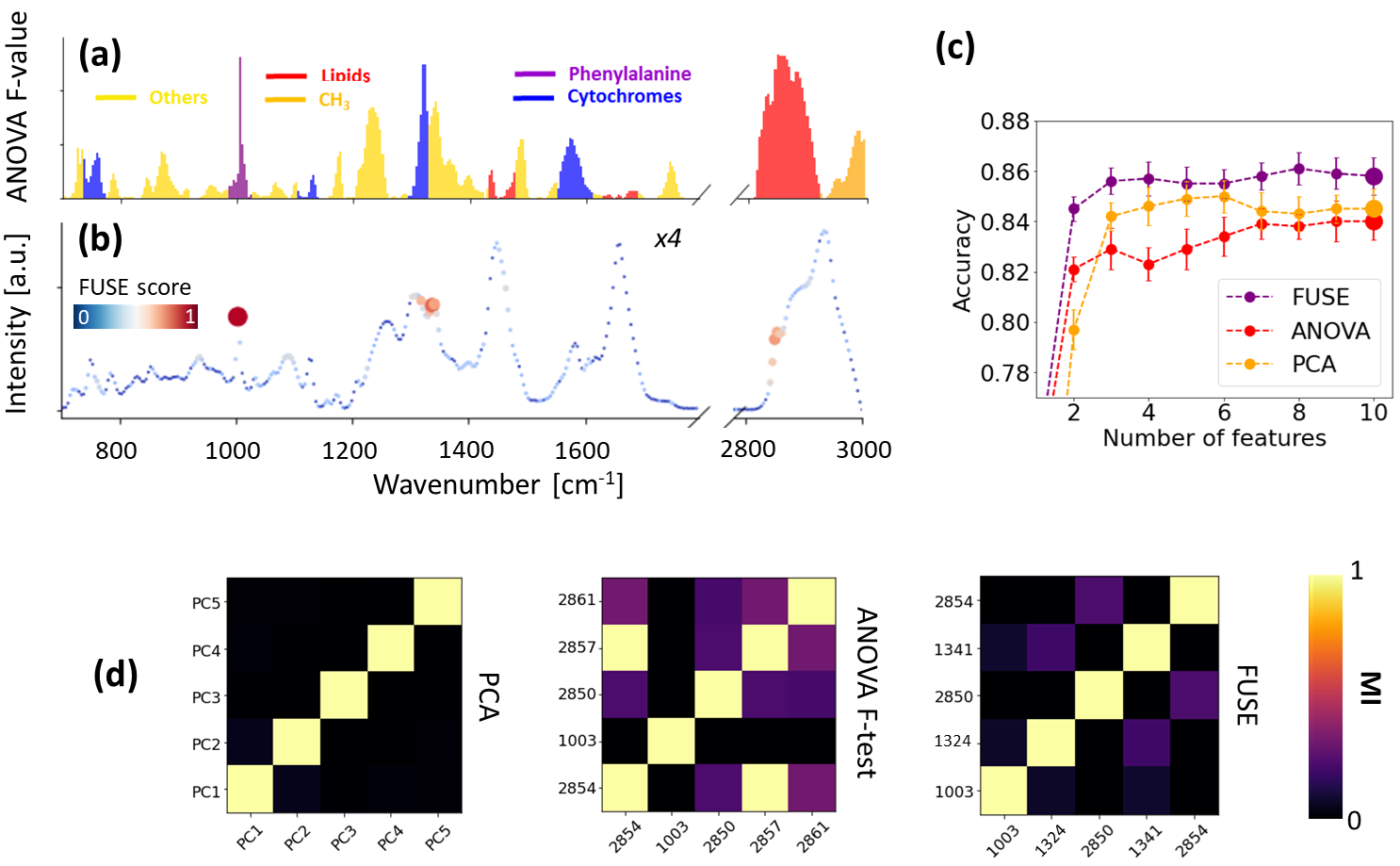}
    \caption{(a) The ANOVA $F$-score of each Raman shift, colored according to their main chemical significance (Table~\ref{table:Ramanshift}). (b) Cytoplasm Raman spectra with wavenumbers color-coded and scaled in proportion to their importance as determined by the FUSE feature selection method, with the intensities of Raman shifts multiplied by a factor of 4 from $\SI{700}{cm^{-1}}$ to $\SI{1800}{cm^{-1}}$ for visual clarity. Top 3 wavenumbers are \SI{1003}{cm^{-1}} (70\%), \SI{1324}{cm^{-1}} (30\%), \SI{2850}{cm^{-1}} (16\%)  (c) Classification results with a 10NN classifier after taking the top \textit{k} features ranked from different dimensionality reduction methods (PCA, ANOVA, FUSE). The accuracy corresponds to 10-fold cross-validation classification accuracy averaged over 1000 independent trials, with the error bar being the standard deviation. (d) Normalized mutual information between the top 5 features ranked after their PC explained variance, ANOVA $F$-score and FUSE score.}
    \label{fig:chemical_signatures}
\end{figure}

\noindent While classification with PCA provides a simple visualization tool to illustrate Raman spectra variability over single cells, it gives limited information regarding the relevant chemical signatures for the diagnosis of FTC. As an alternative interpretable frameworks, we trained a $k$NN classifier on a specific subset of wavenumbers. By selecting the top $N$ wavenumbers that optimize classification accuracy, we can identify the most relevant wavenumbers for FTC diagnosis in cytoplasms.
However, as Raman spectra in our setup span over about 1000 wavenumbers, it is impractical to attempt all of the possible combinations of $N$ wavenumbers. To overcome this, we first applied an \textit{univariate} feature selection approach that scores each feature individually and independently. The best set of wavenumbers is then defined by taking the top ranked $N$ wavenumbers. Here, we ranked features by their ANOVA $F$-test~\cite{box1953non} score, measuring how the variance within FTC and NT class compare to the variance between groups for that specific wavenumber. Figure~\ref{fig:chemical_signatures}a presents the ANOVA $F$-value of Raman intensities over wavenumbers, with colors indicating the molecular components assigned for each wavenumber. Interestingly, a wide range of wavenumbers was evaluated as being important, with five groups of Raman shifts, lipids, CH$_3$, phenylalanine, cytochromes, and others, being relevant for FTC diagnosis. It should be noted that to choose top $N$ relevant wavenumbers is not so straightforwards.

However, this method does not account for potential inter-dependency among the features, which makes it difficult to find the best combination of Raman shifts for the classification, especially since many wavenumbers are highly correlated to each other. Thus, we complemented our analysis with FUSE~\cite{gaudel2010feature}, a more advanced \textit{multivariate} feature selection algorithm that takes into account dependencies among feature variables. Briefly, FUSE formalizes feature selection as a reinforcement learning problem, and relies on Monte Carlo tree search to explore efficiently the feature set space by exploiting the temporally acquired information along the exploration~\cite{gaudel2010feature, pelissier2018feature} to identify the best feature set (Methods Section~\ref{FUSE}). The search was ran independently 10000 times on different subsampling of the dataset, and features consistently selected across multiple searches were given a high FUSE score. Figure~\ref{fig:chemical_signatures}b depicts the FUSE score represented by color grade on the top of the average Raman spectrum of FTC and NT cells. The more the color is reddish, the more the corresponding wavenumber is important for differentiating FTC and NT cells. We find that the relevant wavenumbers are more localized to those associated with lipids, phenylalanine, and cytochromes.


In Figure~\ref{fig:chemical_signatures}c, we highlight the performances of a $10$NN classifier trained with the obtained set of features by the two approaches described above in addition to a set of linear combination of Raman intensities expressed by PCs.  The accuracies corresponds to the 5-fold cross-validation classification accuracy (averaged over 1000 independent sampling). Interestingly, features selected by FUSE outperform other methods by one to two percents (from 84\% to 86\%). FUSE based 10NN classifier yields an accuracy of \SI{86}{\%} using the intensities of only three Raman shifts at \SI{1003}{cm^{-1}}, \SI{1324}{cm^{-1}} and \SI{2850}{cm^{-1}}, for which the peaks may be identified as phenylalanine, cytochromes, and lipids. 
Such optimized set of Raman shifts in differentiating FTC and NT as independent as possible from each other may provide us with a more efficient strategy in accelerating the measurements.  

Lastly, to confirm the dependency among the feature variables selected by the there schemes, we computed the normalized mutual information (MI) among the first five feature variables for each (Methods Section~\ref{MI}). It measures the amount of information obtained about one variable through another variable (Figure~\ref{fig:chemical_signatures}d). The top five feature variables selected by FUSE are more independent of each other (MI<0.1), compared to those by ANOVA (MI up to 0.25). ANOVA takes some redundant features arsing from high wavenumber region that represents lipids. The Mutual information between the PCs is very low (MI<0.1) as expected, since it is a direct result from their mathematical construction. 

\newpage

\section{Discussion}

\noindent Morphological abnormalities of nucleus are key features in cancer cytopathology, and their underlying causes have been attributed to a range of biochemical alterations in nuclear components~\cite{zink2004nuclear, fischer2020nuclear, singh2022nuclear}. As a result, Raman spectral analysis of the cell nucleus has been employed to detect cancer across a range of cytology samples, including cervical, oral, and bladder cells~\cite{duraipandian2018raman, traynor2021raman, o2021automated}. Notably, high-grade precancerous cells have been detected even when their morphology appeared normal, highlighting the capability of Raman micro-spectroscopy to identify biochemical changes prior to observable morphological alterations~\cite{traynor2021raman}. In the context of thyroid cells, Raman spectroscopy was employed to distinguish anaplastic thyroid cancer cell lines from normal thyroid cells by detecting higher intensities of Raman bands associated with DNA-rich components (e.g., O-P-O backbone stretching, nucleic acids, $\alpha$-helix), aligning with the expectation that cancer cells have elevated levels of nuclear material due to increased mitotic activity~\cite{harris2009raman}. In contrast, follicular thyroid carcinoma have been more effectively distinguished by Raman bands corresponding to cytoplasmic components, such as carotenoids, cytochromes, and lipid droplets~\cite{rau2017proof, taylor2019high, sbroscia2020thyroid, hayakawa2023lipid}. While nucleic acid-associated Raman bands have also been highlighted as key features for discriminating follicular thyroid cancers in several studies~\cite{teixeira2009thyroid, rau2017proof}, the precise spatial origin of these signals, whether from the cytoplasm or nucleus, remained unclear.

While prior research on FTC has predominantly focused on analyzing cellular aggregates~\cite{taylor2019high, sbroscia2020thyroid, wang2023differentiating}, here we utilized single-cell Raman imaging of FTC and NT cells to assess the contribution of different cellular regions to FTC, specifically comparing the nucleus and cytoplasm. Interestingly, we observed significant spatial variability in the "carcinogenic areas" within individual cells, with most of the informative signals originating from the cytoplasm, suggesting a lower degree of nuclear biochemical alterations in these cells compared to other cancer types. This observation is consistent with the understanding that follicular thyroid cancer cells are relatively well-differentiated, retaining many features of normal thyroid cells, including nuclear morphology and organization~\cite{asa2019current}. As a result, FTC cells may not exhibit the same degree of nuclear alterations typically associated with more aggressive and less differentiated cancers. This reflects their lower malignancy and reduced propensity for rapid genetic changes, which aligns with their slower progression and more favorable prognosis compared to less differentiated cancers~\cite{sobrinho2011follicular}.

Regarding the biological changes observed in the cytoplasm, feature selection analysis identified an optimal set of nearly independent wavenumbers primarily associated with lipids, cytochromes, and proteins containing phenylalanine moieties. The observed relative increase in lipid concentration in the cytoplasm of FTC cells is consistent with studies on aberrant lipid biosynthesis in various cancers~\cite{menendez2007fatty,baenke2013hooked}, including FTC~\cite{hayakawa2023lipid}, which have led to the identification of new potential drug targets for thyroid cancer~\cite{von2015aberrant}. Similarly, cytochrome c release has been implicated in several therapeutic strategies against thyroid cancer~\cite{bikas2020cytochrome,pan2001cytochrome}, while phenylalanine metabolism has been found to influence the translational processes of RAS mutations~\cite{fagin2004thyroid,m2011diagnostic}, which are known to play critical roles in the MAPK and PI3K signaling pathways in thyroid cancer~\cite{zaballos2017key}.

Our work also addressed a critical issue in Raman analysis: inconsistencies in devices, dates, and images from Raman measurements~\cite{guo2020comparability}, which are a major hindrance to the applications of Raman spectroscopy, especially for samples whose differences in Raman signals are not significantly larger than such uncertainties. By employing our recently developed EBC method~\cite{taylor2023correction}, which leverages the darkest regions of the Raman images to correct for background noise, we ensured that these organelle-focused findings remained robust despite these variations. In fact, despite substantial differences between the two devices used in our study—specifically in terms of objective lens, pixel resolution, laser power, and exposure time (Table~\ref{table:setup}), our spectra processed with EBC ~\cite{taylor2023correction} remained consistent (malignancy prediction accuracy $> \SI{80}{\%}$), even when making prediction by excluding cells measured from the same device (Figure~\ref{fig:summary}d, Table~\ref{table:classif_prot}). Also, our ability to clearly identify the nuclear and cytoplasmic area with Raman spectroscopy (AUC of 0.95) could introduce new diagnostic utilities, aligning with routine practices such as nuclear staining but offering non-destructive and rapid analysis advantages.

In the co-cultured system, where FTC (FTC-133 or RO82W-1) and NT (Nthy-ori 3-1) cells interact with each other, we showed that knowledge acquired from independent FTC and NT cell cultures was transferable to that co-cultured system. With the ground truth assignment based on the injected SV40T fluorescence measurement, we evaluated single cells classification and obtained an accuracy comparable to the one obtained when evaluating independent FTC and NT cell lines. 
Chemical micro-environment composed of solely FTC and NT cells may not always be the same compared to that of co-cultured systems, and interactions between FTC and NT cells might induce the change of states of cells~\cite{ivers2014dynamic}. While extensively used in medical research for their convenience of use, cell lines do not capture tissue complexity and heterogeneity, mainly because they consist of a single cell type that is adapted to grow in culture and lacks interactions with other cell types, and thus differ from tissues environments in several biochemical aspects~\cite{lopes2017regulatory}. As an example, previous literature on hyperspectral Raman images of thyroid tissues~\cite{rau2017proof} emphasized an increased presence of carotenoids in FTC tissues, these were however not directly observed in our measurements with no obvious peaks at carotenoids wavenumbers in our spectra ($\SI{1006}{cm^{-1}}$, $\SI{1156}{cm^{-1}}$, $\SI{1520}{cm^{-1}}$). In this context, a similar Raman imaging analysis should be performed on malignant and normal thyroid tissues rather than cell lines to confirm the validity of our results. While preliminary work in this direction already confirmed the applicability of Raman imaging for the diagnosis of thyroid tissues~\cite{rau2017proof,rau2016raman,rau2019raman}, an extensive Raman characterisation of all thyroid cancer variants, FTC, PTC, MTC and ATC, would be especially desirable for the development of reliable Raman diagnosis tool. 

Finally, as the Raman acquisition of each image in this study span 40 to 90 min, speeding up the measurement could be considered to facilitate clinical application. As only a few Raman bands contribute to cancer detection, the process could be expedited by selectively measuring these cancer-relevant bands~\cite{mochizuki2023high}. Furthermore, our findings also indicate that an almost equivalent AUC (with only a \SI{1}{\%} decrease) can be obtained by considering just one quarter of randomly sampled pixels in the cytoplasm area. This suggests that cell malignancy can be accurately predicted without measuring the entire cell, allowing the diagnosis to be conducted without the need for complete image acquisition~\cite{zhang2018dynamic}. The prediction confidence would increase as more of cytoplasmic area is measured, so the acquisition could be ended once the desired confidence threshold is reached~\cite{tabata2020bad, tabata2024fly}. Such partial measurements, whether spatial or spectral, are made possible through Raman microscopy with programmable multipoints illumination~\cite{qi2012parallel,qi2013high, mochizuki2023high} as an alternative to the line scanning procedure used in our measurements. Nevertheless, an additional framework is required to reliably separate non-cellular, cytoplasmic and nucleus regions in the Raman image, because such separation was performed after full acquisition of the Raman image in the current work, and is less reliable with partial image information. In this context, supporting Raman imaging with fluorescence measurements is a viable solution to identify cell's cytoplasm and nucleus~\cite{uzunbajakava2003combined,bennet2014simultaneous}. Still, such simultaneous Raman-fluorescence experimental setup is constraining, and an automated cellular segmentation pipeline from stacked microscope images~\cite{selinummi2009bright} would offer a more scalable and cheaper alternative.

\section{Materials and Methods}

\subsection{Sample preparation}
\label{sample_prep}
\noindent Human thyroid follicular carcinoma cell lines (FTC-133 and RO82W-1) and human thyroid follicular epitherial cell line (Nthy-ori 3-1) were employed for this research as cancer and non-cancer cells, respectively. The cells were seeded on gelatin-coated dish with a calcium fluoride substrates window (CRYSTRAN LTD, Raman Grade $\text{CaF}_2$ CAFP13-0.2). The cell density was \SI{3.0e5}{} cells/ dish. As for the culturing media, DMEM/Ham’s F-12 (FUJIFILM Wako Pure Chemical Corporation, 61-23201) and RPMI1640 (NACALAITESQUE, INC., 05176-25) were used as basal media for follicular carcinoma cell lines and follicular epitherial cell line, respectively. The basal media were supplemented with 10\% fetal bovine serum (FBS) (GE Healthcare, SH3-920.03) and 1\% penicillin-streptomycin-L- glutamine solution (FUJIFILM Wako Pure Chemical). The cells were cultured in an incubator at \SI{37}{\degree C} with 5\% $\text{CO}_\text{2}$ and saturated humidity. After 40 to 48 hours incubation, the cells were fixed. The culture media was removed and the cells were washed with $\text{Ca}^{2+}$ and $\text{Mg}^{2+}$+ free Phosphate Buffered Saline (PBS(-)). Then the cells were treated with 4\% paraformaldehyde phosphate buffer solution for 10 min. After the treatment, the cells were washed with PBS(-) three times for 5 min each time. Prior to Raman imaging, PBS(-) was filled in culture dish to prevent the drying of the cells.

As for the co-culture system, cancer cell line (FTC-133, RO82W-1) and non-cancer cell line (Nthy-ori 3-1) were cultured together in the cell density of \SI{0.5e5}{} cells/ dish each. RPMI1640 supplemented with 10\% FBS and 1\% penicillin-streptomycin-L-glutamine solution was employed for culture medium. The cell fixation procedure was the same as the one used for independent cell-line culturing.

\subsection{Immunofluorescence imaging for nucleus identification and FTC/NT identification in the co-culture system}
\label{immaging}

\noindent For the identification nucleus regions, and the the discrimination of cancer (FTC-133 and RO82W-1) or non-cancer (Nthy-ori 3-1) cell lines in mixed culture, immunofluorescent study was performed after Raman microscopy. Nuclear Staining was performed with TO-PRO-3 iodide, and SV40 large T protein was targeted as a marker for identification of Nthy-ori 3-1 cell line. 

After Raman microscopy, the cells were treated with \SI{0.2}{\%} Triton-X in PBS(-) for 1 min to permeabilize their membrane, then blocked with \SI{10}{\%} FBS, \SI{0.1}{\%} Triton-X in PBS(-) at \SI{37}{\degree C} for 1h. Mouse monoclonal anti-SV40 large T protein antibody (at a 1/50 dilution, abcam, ab16879) was treated at \SI{37}{\degree C} for 2h. After the primary antibody treatment, the samples were washed with PBS(-) three times for 5 min each. Alexa Fluor 488 conjugated to anti-mouse IgG (at a 1/100 dilution, invitrogen, A11029) was treated in the dark for 1 h at room temperature. The cell nuclei were stained with TO-PRO-3 iodide (at a 1/500 dilution, invitrogen, T3605) for 30 min after the secondary antibody treatment at room temperature. Then the samples were rinsed with PBS(-) tree times each for 5 min. The images were acquired with a confocal laser scanning microscope (FV1000, Olympus). Alexa Fluor 488 and TO-PRO-3 were excited at 488 nm (10\% power) and \SI{633}{nm} (10\% power), respectively. The emission light was collected with a x20 objective lens (N.A. 0.75, UPLSAPO, Olympus). The emission wavelengths were \SI{520}{nm} and \SI{664}{nm} for Alexa fluor 488 and TO-PRO-3, respectively.

\subsection{Data acquisition and Raman Microscopy}

\label{data_acquisition}

\noindent Raman imaging of cells was performed by a line-illumination Raman microscopy~\cite{veirs1990mapping,hamada2008raman}. This line-illumination scheme enables the parallel detection of Raman spectra from multiple points in the sample in each exposure, resulting in the acceleration of image acquisition rate, typically 2 order of magnitudes faster than conventional point-illumination Raman microscope~\cite{palonpon2013raman}. As a result, the acquisition time of a single Raman image ranged from 40 minutes to 60 minutes in our experiment.

The excitation laser light was shaped into line pattern by a cylindrical lens, then focused on the sample through an objective lens. The Raman scattering induced along the illuminated laser line on the sample was collected with the same objective lens, and then relayed to the slit entrance of a spectrophotometer. On the relaying way, the remaining Rayleigh scattering light was eliminated by longwave-pass edge filters. Raman scattering passing through the entrance slit, which corresponds to the Raman scattering induced by the laser line focus on the sample due to slit-confocal effect, was detected by a cooled CCD camera after being dispersed by a grating inside the spectrophotometer. The output CCD image in a single exposure provided a spectral image, for which the $y$-axis corresponds to the spatial distribution along the illuminated line and $x$-axis the spectral frequency (wavenumber). During the imaging process, this CCD exposure was repeated along the direction perpendicular to the line illumination, and the scanning of laser line was manipulated by a galvanometer mirror so that the location of the illuminated laser line on the sample and the entrance slit remained conjugated. The Raman spectral dataset obtained through the imaging process consists of 3-dimensional $(x, y, \lambda)$ information as generally-called hyperspectral images.

In this research, we used two different devices, i.e., commercial device system (\textit{Nanophoton}, RAMAN-11) to conduct Raman Microscopy (termed device 1), and a home-built device system (device 2). Although both device systems adopted a \SI{532}{nm} CW laser as the excitation light source, the subsequent measurement conditions and installed devices were different in several aspects, as described in Supplementary Table \ref{table:setup}.

\subsubsection*{Raman Data calibration}

\label{data_calibration}

\noindent Data calibration consists of assigning the spectral axis of hyperspectral Raman images to their corresponding wavenumber values~\cite{kumamoto2019high}. An ethanol spectrum was measured prior or posterior to Raman measurements on each date, and the seven major bands of ethanol (884, 1052, 1096, 1454, 2880, 2930, and \SI{2974}{cm^{-1}}) were used as reference bands for calibration. Generation of the wavenumber axis from the original axis was performed by fitting the seven points with a third order polynomial function.

\subsection{Raman Data Preprocessing}

\label{data_processing}

\noindent Raman data typically contain substantial noise and background due to fluorescence, water, the substrate used, and cosmic rays. Standard preprocessing methods used in our work involve cosmic ray removal, denoising by singular value decomposition, background correction by polynomial fitting protocol, and area-normalization. Raman images were first preprocessed to remove electronic bias, cosmic rays, and correct for irradiance inhomogeneity. Singular-value decomposition was used to increase signal-to-noise ratio, with the 7 most significant singular values being retained during reconstruction. Baseline correction was performed with a $7^{\text{th}}$ degree recursive polynomial fitting algorithm (Polyfit)~\cite{lieber2003automated} over the entire spectral range (700-$\SI{3000}{cm^{-1}}$). Intensities in the silent region (1800-2800 cm$^{–1}$) were set to zero before area-normalizing the sum of intensities over the remaining wavenumbers (700-$\SI{1800}{cm^{-1}}$) and (2800-$\SI{3000}{cm^{-1}}$) to unity. For further increasing the signal-to-noise ratio, rather than using a simple binning scheme on a square grid of predetermined size and location, adjacent pixels were grouped over spatially-local regions as superpixels~\cite{achanta2012slic}, whose Raman intensities are averaged over the adjacent pixels while maintaining the spatial structure of Raman images. The obtained superpixels contained an average of 10 individual pixels, corresponding to an average area of $\sim 1\mu \mathrm{m}^{2}$. Refer to Supplementary Section~\ref{sup_method} for full details of the data preprocessing procedures. The code utilized for processing Raman images can be accessed publicly on GitHub via \url{https://github.com/AI-SysBio/Raman-Imaging-Processing}.

\subsection{Identification of cellular structures}  

\label{method:cell_identification}

\subsubsection*{Identification of background and cell regions with EBC}

\noindent The buffer solution used in our measurements is phosphate buffered saline, characterized by a weak Raman contribution in the high wavenumber region (2800-$\SI{3040}{cm^{-1}}$). Identification of background and cell regions is based on clustering in this wavenumber range.

The identification of background and cell regions was done on raw images, and performed by $k$-means clustering of the Raman intensity averaged over the high wavenumber region (2800-$\SI{3040}{cm^{-1}}$). Each Raman spectra image was first partitioned into $n$ groups, and the spectra belonging to the least intense cluster were retained as pixels containing background regions. The choice of the number of group $n$ is relative to each Raman image because it depends on the cell density within the image. To ensure the consistency of background subtraction through different images, the minimum number of group $n$ was chosen such as the average distance within each cluster was smaller than the distance produced by Poisson error. Details are provided in Supplementary Section~\ref{sup_method}. Then, the cellular regions were assigned as being composed of 
the $m$ most intense clusters and the value of $m$ was chosen to maximize the accuracy of FTC/NT classification. It was found that, at least for $2\le n \le 20$, $m$ tends to be about half of $n$ (See Supplementary Section~\ref{sup:mclusters}). 



\subsubsection*{Cell isolation}
\noindent A binary mask was produced from the previous identification of cellular region, where a pixel is associated to one if it is from a cellular region, and zero otherwise. The connected regions of ones, containing cells, were identified with a flood-fill algorithm~\cite{nosal2008flood}. Isolated regions containing less than 200 pixels ($< \sim 200 \mu \mathrm{m}^{2}$) are too small to be cells and were discarded. In some Raman images, some strong overlap exists between the cells because of high cell density. In these cases, cells were segmented manually.

\subsection{Identification of nucleus and cytoplasm regions}
\label{method:nucleus_identification}
\noindent The identification of nucleus and cytoplasm regions was performed on the cellular regions assigned in Method \ref{method:cell_identification}. Differentiation of nucleus region relative to cytoplasm region was performed by $k$-means clustering on the average intensity of $\text{CH}_3$ stretching region 2930-$\SI{3000}{cm^{-1}}$. We found that clustering the cellular regions into 8 groups and keeping the 5 largest intensity as cytoplasm region and those having the remaining 3 groups as nuclei was maximizing the F1 score and Matthews correlation coefficient (MCC)~\cite{chicco2020advantages} score to the ground truth on our 10 stained images (Supplementary Figure~\ref{fig:ground_nuc_SI}c,d). Note that here, since the variation in cell density across images was already considered in the previous steps (background subtraction and cellular identification), the same number of clusters can be used for all images.

\subsection{Feature selection with FUSE}

\label{FUSE}

\noindent Feature Uct (Upper confidence bounds applied to trees)  SElection (FUSE)\cite{gaudel2010feature} is a \textit{multivariate} feature selection algorithm aiming at selecting the optimal feature set for a specific task, taking into account dependencies among feature variables. As it is impractical to evaluate all of the possible combinations of feature set when the number of feature except when the number of features is very low, FUSE formalizes feature selection as a reinforcement learning problem, and relies on Monte Carlo tree search to explore efficiently the feature set space by exploiting the temporally acquired information along the exploration to identify the best feature set. Here, The feature set space is represented as a Directed Acyclic Graph, and feature subsets are evaluated with a $k$NN classifier with $k=10$.

Importantly, FUSE does not compute a score for each feature, but rather returns the best combination of features for a given dataset. As the algorithm is not deterministic, we ran FUSE independently 10000 times and we defined a score for each features as the proportion of runs it was included in the best feature set after 10000 search iterations. To prevent overfitting, we ran FUSE on 75\% of the original dataset and reshuffled the dataset each time. The \texttt{C++} implementation from~\cite{pelissier2018feature} was used.

\subsection{Mutual information}

\label{MI}

\noindent In Shanon's Information theory, to quantify ``amount of information'' obtained about one random variable $X$ through the other random variable $Y$, one can calculate the Mutual Information (MI) between the two stochastic variables, defined as follows:
\begin{equation}
    \nonumber
    \textit{MI}\hspace{2pt}(X,Y) = \mathlarger{\mathlarger{\sum}}_{x \in X} \ \mathlarger{\mathlarger{\sum}}_{y \in Y} \ p(x,y) \log \frac{p(x,y)}{p(x)p(y)} 
\end{equation}
where $p(x,y)$ is the joint probability density function of $X$ and $Y$, and $p(x)$ and $p(y)$ are the marginal probability distribution functions of $X$ and $Y$, respectively. The mutual information for target variables is estimated with a nonparametric methods based on entropy estimation from $k$-nearest neighbors distances~\cite{kraskov2004estimating} implemented in \texttt{Python} (\texttt{sklearn} library). To ensure consistency across various sample size, the mutual information was normalised by the maximum mutual information for these samples (i.e.,$\min\{H(X),H(Y)\}$ where, e.g., $H(X)$ is the Shannon entropy).

\section*{Data and code Availability}
\noindent All spectra measured from both device~1 (Commercial) and device 2 (Home-Built) analyzed during the current study are available on Mendeley~\cite{kyoto2019Raman, osaka2019Raman}. Our \texttt{python} implementation for the processing methods used to analyze Raman hyperspectral images is publicly available on Github at \url{https://github.com/AI-SysBio/Raman-Imaging-Processing}.

\subsection*{Acknowledgements and Funding}
\noindent This research was supported by JSPS, Grant-in-Aid for Scientific Research (No. 25287105) (to T.K.), Grant-in-Aid for Exploratory Research (No. 25650044) (to T.K.), Grant-in Aid for Scientific Research on Innovative Areas (Singularity biology) (No. 18H05408) and (Chemistry for Multi-Molecular Crowding Biosystems) (No. 18H04530) (to T.K.), Grant-in-Aid for Early-Career Scientists (No. 20K15195)(to K.M.), the Imaging Science Project of the Center for Novel Science Initiatives (CNSI), National Institutes of Natural Sciences (NINS) (Grant Number IS281002) (to T.K.). Japan Science and Technology Agency (JST) / Core Research for Evolutional Science and Technology (CREST), Grant Number JPMJCR1662, Japan (to T.K., K.F., Y.H.), and the Research Program of ”Dynamic Alliance for Open Innovation Bridging Human, Environment and Materials” in ”Network Joint Research Center for Materials and Devices” (to K.F., Y.H.). Institute for Chemical Reaction Design and Discovery (ICReDD) was established by World Premier International Research Initiative (WPI), MEXT, Japan.

\subsection*{Author contributions statement}
\noindent A.P. analyzed the measurements and wrote the first draft of the manuscript, under the supervision of T.K. The experimental design including the choice of FTC as the target, choice of the cell lines, and the use of the co-culturing system, was developed by Y.H. and Y.K, and executed by K.H. and KM.. K.H. prepared the cell lines, conducted device~1 and co-culture measurements, K.M. conducted device 2 measurements under the supervision of K.F., Y.K. established the wavenumber calibration protocol. Methods section were written by A.P., K.H., K.M. and  N.T. All authors participated in the regular discussions and revised the manuscript.

\subsection*{Competing Interests}

\noindent The authors declare no competing interests.\\

\printbibliography[title={Bibliography}]

@article{taylor2019high,
  title={High-Resolution Raman Microscopic Detection of Follicular Thyroid Cancer Cells with Unsupervised Machine Learning},
  author={Taylor, J Nicholas and Mochizuki, Kentaro and Hashimoto, Kosuke and Kumamoto, Yasuaki and Harada, Yoshinori and Fujita, Katsumasa and Komatsuzaki, Tamiki},
  journal={The Journal of Physical Chemistry B},
  year={2019},
  publisher={ACS Publications}
}

@article{santos2017raman,
  title={Raman spectroscopy for cancer detection and cancer surgery guidance: translation to the clinics},
  author={Santos, In{\^e}s P and Barroso, Elisa M and Schut, Tom C Bakker and Caspers, Peter J and van Lanschot, Cornelia GF and Choi, Da-Hye and van der Kamp, Martine F and Smits, Roeland WH and van Doorn, Remco and Verdijk, Rob M and others},
  journal={Analyst},
  volume={142},
  number={17},
  pages={3025--3047},
  year={2017},
  publisher={Royal Society of Chemistry}
}

@article{duraiyan2012applications,
  title={Applications of immunohistochemistry},
  author={Duraiyan, Jeyapradha and Govindarajan, Rajeshwar and Kaliyappan, Karunakaran and Palanisamy, Murugesan},
  journal={Journal of pharmacy \& bioallied sciences},
  volume={4},
  number={Suppl 2},
  pages={S307},
  year={2012},
  publisher={Wolters Kluwer--Medknow Publications}
}

@article{khan2001classification,
  title={Classification and diagnostic prediction of cancers using gene expression profiling and artificial neural networks},
  author={Khan, Javed and Wei, Jun S and Ringner, Markus and Saal, Lao H and Ladanyi, Marc and Westermann, Frank and Berthold, Frank and Schwab, Manfred and Antonescu, Cristina R and Peterson, Carsten and others},
  journal={Nature medicine},
  volume={7},
  number={6},
  pages={673},
  year={2001},
  publisher={Nature Publishing Group}
}

@article{ikeda2018raman,
  title={Raman spectroscopy for the diagnosis of unlabeled and unstained histopathological tissue specimens},
  author={Ikeda, Haruo and Ito, Hiroaki and Hikita, Muneaki and Yamaguchi, Noriko and Uragami, Naoyuki and Yokoyama, Noboru and Hirota, Yuko and Kushima, Miki and Ajioka, Yoichi and Inoue, Haruhiro},
  journal={World journal of gastrointestinal oncology},
  volume={10},
  number={11},
  pages={439},
  year={2018},
  publisher={Baishideng Publishing Group Inc}
}

@article{cui2018raman,
  title={Raman Spectroscopy and Imaging for Cancer Diagnosis},
  author={Cui, Sishan and Zhang, Shuo and Yue, Shuhua},
  journal={Journal of Healthcare Engineering},
  volume={2018},
  year={2018},
  publisher={Hindawi}
}

@article{rau2017proof,
  title={Proof-of-concept Raman spectroscopy study aimed to differentiate thyroid follicular patterned lesions},
  author={Rau, Julietta V and Fosca, Marco and Graziani, Valerio and Taffon, Chiara and Rocchia, Massimiliano and Caricato, Marco and Pozzilli, Paolo and Muda, Andrea Onetti and Crescenzi, Anna},
  journal={Scientific Reports},
  volume={7},
  number={1},
  pages={14970},
  year={2017},
  publisher={Nature Publishing Group}
}

@article{rau2016raman,
  title={RAMAN spectroscopy imaging improves the diagnosis of papillary thyroid carcinoma},
  author={Rau, Julietta V and Graziani, Valerio and Fosca, Marco and Taffon, Chiara and Rocchia, Massimiliano and Crucitti, Pierfilippo and Pozzilli, Paolo and Muda, Andrea Onetti and Caricato, Marco and Crescenzi, Anna},
  journal={Scientific reports},
  volume={6},
  pages={35117},
  year={2016},
  publisher={Nature Publishing Group}
}

@article{wang2010detection,
  title={Detection and classification of thyroid follicular lesions based on nuclear structure from histopathology images},
  author={Wang, Wei and Ozolek, John A and Rohde, Gustavo K},
  journal={Cytometry Part A: The Journal of the International Society for Advancement of Cytometry},
  volume={77},
  number={5},
  pages={485--494},
  year={2010},
  publisher={Wiley Online Library}
}

@article{butler2016using,
  title={Using Raman spectroscopy to characterize biological materials},
  author={Butler, Holly J and Ashton, Lorna and Bird, Benjamin and Cinque, Gianfelice and Curtis, Kelly and Dorney, Jennifer and Esmonde-White, Karen and Fullwood, Nigel J and Gardner, Benjamin and Martin-Hirsch, Pierre L and others},
  journal={Nature protocols},
  volume={11},
  number={4},
  pages={664},
  year={2016},
  publisher={Nature Publishing Group}
}

@article{sobrinho2011follicular,
  title={Follicular thyroid carcinoma},
  author={Sobrinho-Simoes, Manuel and Eloy, Catarina and Magalhaes, Joao and Lobo, Cl{\'a}udia and Amaro, Teresina},
  journal={Modern Pathology},
  volume={24},
  number={S2},
  pages={S10},
  year={2011},
  publisher={Nature Publishing Group}
}

@article{leboulleux2016papillary,
  title={Papillary thyroid microcarcinoma: time to shift from surgery to active surveillance?},
  author={Leboulleux, Sophie and Tuttle, R Michael and Pacini, Furio and Schlumberger, Martin},
  journal={The lancet Diabetes \& endocrinology},
  volume={4},
  number={11},
  pages={933--942},
  year={2016},
  publisher={Elsevier}
}

@article{jegerlehner2017overdiagnosis,
  title={Overdiagnosis and overtreatment of thyroid cancer: a population-based temporal trend study},
  author={Jegerlehner, Sabrina and Bulliard, Jean-Luc and Aujesky, Drahomir and Rodondi, Nicolas and Germann, Simon and Konzelmann, Isabelle and Chiolero, Arnaud and NICER Working Group and others},
  journal={PloS one},
  volume={12},
  number={6},
  pages={e0179387},
  year={2017},
  publisher={Public Library of Science}
}

@article{palonpon2013raman,
  title={Raman and SERS microscopy for molecular imaging of live cells},
  author={Palonpon, Almar F and Ando, Jun and Yamakoshi, Hiroyuki and Dodo, Kosuke and Sodeoka, Mikiko and Kawata, Satoshi and Fujita, Katsumasa},
  journal={Nature protocols},
  volume={8},
  number={4},
  pages={677},
  year={2013},
  publisher={Nature Publishing Group}
}

@article{pence2016clinical,
  title={Clinical instrumentation and applications of Raman spectroscopy},
  author={Pence, Isaac and Mahadevan-Jansen, Anita},
  journal={Chemical Society Reviews},
  volume={45},
  number={7},
  pages={1958--1979},
  year={2016},
  publisher={Royal Society of Chemistry}
}

@article{hamada2008raman,
  title={Raman microscopy for dynamic molecular imaging of living cells},
  author={Hamada, Keisaku and Fujita, Katsumasa and Smith, Nicholas Isaac and Kobayashi, Minoru and Inouye, Yasushi and Kawata, Satoshi},
  journal={Journal of biomedical optics},
  volume={13},
  number={4},
  pages={044027},
  year={2008},
  publisher={International Society for Optics and Photonics}
}

@article{von2015aberrant,
  title={Aberrant lipid metabolism in anaplastic thyroid carcinoma reveals stearoyl CoA desaturase 1 as a novel therapeutic target},
  author={Von Roemeling, Christina A and Marlow, Laura A and Pinkerton, Anthony B and Crist, Angela and Miller, James and Tun, Han W and Smallridge, Robert C and Copland, John A},
  journal={The Journal of Clinical Endocrinology \& Metabolism},
  volume={100},
  number={5},
  pages={E697--E709},
  year={2015},
  publisher={Oxford University Press}
}

@article{suen2002fine,
  title={Fine-needle aspiration biopsy of the thyroid},
  author={Suen, Kenneth C},
  journal={Canadian Medical Association Journal},
  volume={167},
  number={5},
  pages={491--495},
  year={2002},
  publisher={Can Med Assoc}
}

@article{dean2015fine,
  title={Fine-needle aspiration biopsy of the thyroid gland},
  author={Dean, Diana S and Gharib, Hossein},
  year={2015},
  journal={MDText. com, Inc.}
}

@article{cibas2009bethesda,
  title={The Bethesda system for reporting thyroid cytopathology},
  author={Cibas, Edmund S and Ali, Syed Z},
  journal={American journal of clinical pathology},
  volume={132},
  number={5},
  pages={658--665},
  year={2009},
  publisher={Oxford University Press}
}

@article{lieber2003automated,
  title={Automated method for subtraction of fluorescence from biological Raman spectra},
  author={Lieber, Chad A and Mahadevan-Jansen, Anita},
  journal={Applied spectroscopy},
  volume={57},
  number={11},
  pages={1363--1367},
  year={2003},
  publisher={Society for Applied Spectroscopy}
}

@inproceedings{lones2010discriminating,
  title={Discriminating normal and cancerous thyroid cell lines using implicit context representation cartesian genetic programming},
  author={Lones, Michael A and Smith, Stephen L and Harris, Andrew T and High, Alec S and Fisher, Sheila E and Smith, D Alastair and Kirkham, Jennifer},
  booktitle={Evolutionary Computation (CEC), 2010 IEEE Congress on},
  pages={1--6},
  year={2010},
  organization={IEEE}
}

@article{teixeira2009thyroid,
  title={Thyroid tissue analysis through Raman spectroscopy},
  author={Teixeira, Caroline SB and Bitar, Renata A and Martinho, Herculano S and Santos, Andr{\'e} BO and Kulcsar, Marco AV and Friguglietti, Celso UM and da Costa, Ricardo B and Arisawa, Emilia {\^A}L and Martin, Airton A},
  journal={Analyst},
  volume={134},
  number={11},
  pages={2361--2370},
  year={2009},
  publisher={Royal Society of Chemistry}
}

@inproceedings{gaudel2010feature,
  title={Feature selection as a one-player game},
  author={Gaudel, Romaric and Sebag, Michele},
  booktitle={International Conference on Machine Learning},
  pages={359--366},
  year={2010}
}

@inproceedings{pelissier2018feature,
  title={Feature selection as Monte-Carlo Search in Growing Single Rooted Directed Acyclic Graph by Best Leaf Identification},
  author={Pelissier, Aurelien and Nakamura, Atsuyoshi and Tabata, Koji},
  booktitle={Proceedings of the 2019 SIAM International Conference on Data Mining},
  pages={450-458},
  year={2019}
}

@article{baenke2013hooked,
  title={Hooked on fat: the role of lipid synthesis in cancer metabolism and tumour development},
  author={Baenke, Franziska and Peck, Barrie and Miess, Heike and Schulze, Almut},
  journal={Disease models \& mechanisms},
  volume={6},
  number={6},
  pages={1353--1363},
  year={2013},
  publisher={The Company of Biologists Ltd}
}

@article{o2018raman,
  title={Raman spectroscopy for the preoperative diagnosis of thyroid cancer and its subtypes: An in vitro proof-of-concept study},
  author={O'dea, Declan and Bongiovanni, Massimo and Sykiotis, Gerasimos P and Ziros, Panos G and Meade, Aidan D and Lyng, Fiona M and Malkin, Alison},
  journal={Cytopathology},
  year={2018},
  publisher={Wiley Online Library}
}

@article{okada2012label,
  title={Label-free Raman observation of cytochrome c dynamics during apoptosis},
  author={Okada, Masaya and Smith, Nicholas Isaac and Palonpon, Almar Flotildes and Endo, Hiromi and Kawata, Satoshi and Sodeoka, Mikiko and Fujita, Katsumasa},
  journal={Proceedings of the National Academy of Sciences},
  volume={109},
  number={1},
  pages={28--32},
  year={2012},
  publisher={National Acad Sciences}
}

@article{czamara2015raman,
  title={Raman spectroscopy of lipids: a review},
  author={Czamara, Krzysztof and Majzner, Katarzyna and Pacia, Marta Z and Kochan, K and Kaczor, A and Baranska, M},
  journal={Journal of Raman Spectroscopy},
  volume={46},
  number={1},
  pages={4--20},
  year={2015},
  publisher={Wiley Online Library}
}

@inproceedings{rubner1998metric,
  title={A metric for distributions with applications to image databases},
  author={Rubner, Yossi and Tomasi, Carlo and Guibas, Leonidas J},
  booktitle={Sixth International Conference on Computer Vision (IEEE Cat. No. 98CH36271)},
  pages={59--66},
  year={1998},
  organization={IEEE}
}

@article{rygula2013raman,
  title={Raman spectroscopy of proteins: a review},
  author={Rygula, A and Majzner, Katarzyna and Marzec, Katarzyna M and Kaczor, Agnieszka and Pilarczyk, Marta and Baranska, M},
  journal={Journal of Raman Spectroscopy},
  volume={44},
  number={8},
  pages={1061--1076},
  year={2013},
  publisher={Wiley Online Library}
}

@article{de2007reference,
  title={Reference database of Raman spectra of biological molecules},
  author={De Gelder, Joke and De Gussem, Kris and Vandenabeele, Peter and Moens, Luc},
  journal={Journal of Raman Spectroscopy: An International Journal for Original Work in all Aspects of Raman Spectroscopy, Including Higher Order Processes, and also Brillouin and Rayleigh Scattering},
  volume={38},
  number={9},
  pages={1133--1147},
  year={2007},
  publisher={Wiley Online Library}
}

@article{kraskov2004estimating,
  title={Estimating mutual information},
  author={Kraskov, Alexander and St{\"o}gbauer, Harald and Grassberger, Peter},
  journal={Physical review E},
  volume={69},
  number={6},
  pages={066138},
  year={2004},
  publisher={APS}
}

@article{pandya2016immune,
  title={The immune system in cancer pathogenesis: potential therapeutic approaches},
  author={Pandya, Pankita H and Murray, Mary E and Pollok, Karen E and Renbarger, Jamie L},
  journal={Journal of immunology research},
  volume={2016},
  year={2016},
  publisher={Hindawi}
}

@article{keeting1992development,
  title={Development and characterization of a rapidly proliferating, well-differentiated cell line derived from normal adult human osteoblast-like cells transfected with SV40 large T antigen},
  author={Keeting, Philip E and Scott, Robert E and Colvard, Douglas S and Anderson, Marlys A and Oursler, Merry J and Spelsberg, Thomas C and Riggs, Lawrence B},
  journal={Journal of Bone and Mineral Research},
  volume={7},
  number={2},
  pages={127--136},
  year={1992},
  publisher={Wiley Online Library}
}

@article{achanta2012slic,
  title={SLIC superpixels compared to state-of-the-art superpixel methods},
  author={Achanta, Radhakrishna and Shaji, Appu and Smith, Kevin and Lucchi, Aurelien and Fua, Pascal and S{\"u}sstrunk, Sabine},
  journal={IEEE transactions on pattern analysis and machine intelligence},
  volume={34},
  number={11},
  pages={2274--2282},
  year={2012},
  publisher={IEEE}
}

@article{kumamoto2019high,
  title={High-Throughput Cell Imaging and Classification by Narrowband and Low-Spectral-Resolution Raman Microscopy},
  author={Kumamoto, Yasuaki and Mochizuki, Kentaro and Hashimoto, Kosuke and Harada, Yoshinori and Tanaka, Hideo and Fujita, Katsumasa},
  journal={The Journal of Physical Chemistry B},
  year={2019},
  publisher={ACS Publications}
}

@article{veirs1990mapping,
  title={Mapping materials properties with Raman spectroscopy utilizing a 2-D detector},
  author={Veirs, D Kirk and Ager, Joel W and Loucks, Eric T and Rosenblatt, Gerd M},
  journal={Applied optics},
  volume={29},
  number={33},
  pages={4969--4980},
  year={1990},
  publisher={Optical Society of America}
}

@inproceedings{nosal2008flood,
  title={Flood-fill algorithms used for passive acoustic detection and tracking},
  author={Nosal, Eva-Marie},
  booktitle={2008 New Trends for Environmental Monitoring Using Passive Systems},
  pages={1--5},
  year={2008},
  organization={IEEE}
}

@article{box1953non,
  title={Non-normality and tests on variances},
  author={Box, George EP},
  journal={Biometrika},
  volume={40},
  number={3/4},
  pages={318--335},
  year={1953},
  publisher={JSTOR}
}

@misc{osaka2019Raman,
  title = {High Resolution Hyperspectral Raman images from Nthy3-1, FTC133 and RO82W-1 cell lines},
  author = {Mochizuki, Kentaro and Hashimoto, Kosuke and Kumamoto, Yasuaki and Pelissier, Aurelien and Harada, Yoshinori and Fujita, Katsumasa},
  doi = {10.17632/dshgffwykw.1},
  year={2019},
  howpublished= {\url{https://data.mendeley.com/datasets/dshgffwykw/1}}
}

@misc{kyoto2019Raman,
  title = {High Resolution Hyperspectral Raman images from Nthy3-1, FTC133 and RO82W-1 cell lines - Commercial device system (Nanophoton, RAMAN-11)},
  author = {Hashimoto, Kosuke and Mochizuki, Kentaro and Kumamoto, Yasuaki and Pelissier, Aurelien and Fujita, Katsumasa and Harada, Yoshinori},
  doi = {10.17632/yz6rvx3zvt.1},
  year={2021},
  howpublished= {\url{https://data.mendeley.com/datasets/yz6rvx3zvt/1}}
}

@article{powers2011evaluation,
  title={Evaluation: from precision, recall and F-measure to ROC, informedness, markedness and correlation},
  author={Powers, David Martin},
  journal={Journal of Machine Learning Technologies },
  volume={2},
  pages={37--63},
  year={2011},
  publisher={Bioinfo Publications}
}

@article{guo2020comparability,
  title={Comparability of Raman spectroscopic configurations: A large scale cross-laboratory study},
  author={Guo, Shuxia and Beleites, Claudia and Neugebauer, Ute and Abalde-Cela, Sara and Afseth, Nils Kristian and Alsamad, Fatima and Anand, Suresh and Araujo-Andrade, Cuauhtemoc and Askrabic, Sonja and Avci, Ertug and others},
  journal={Analytical Chemistry},
  volume={92},
  number={24},
  pages={15745--15756},
  year={2020},
  publisher={ACS Publications}
}

@article{guo2018extended,
  title={Extended multiplicative signal correction based model transfer for Raman spectroscopy in biological applications},
  author={Guo, Shuxia and Kohler, Achim and Zimmermann, Boris and Heinke, Ralf and Stöckel, Stephan and Rösch, Petra and Popp, Jürgen and Bocklitz, Thomas},
  journal={Analytical chemistry},
  volume={90},
  number={16},
  pages={9787--9795},
  year={2018},
  publisher={ACS Publications}
}

@article{taylor2023correction,
  title={Correction for Extrinsic Background in Raman Hyperspectral Images},
  author={Taylor, J Nicholas and P{\'e}lissier, Aur{\'e}lien and Mochizuki, Kentaro and Hashimoto, Kosuke and Kumamoto, Yasuaki and Harada, Yoshinori and Fujita, Katsumasa and Bocklitz, Thomas and Komatsuzaki, Tamiki},
  journal={Analytical Chemistry},
  year={2023},
  publisher={ACS Publications}
}

@article{zaballos2017key,
  title={Key signaling pathways in thyroid cancer},
  author={Zaballos, Miguel A and Santisteban, Pilar},
  journal={Journal of Endocrinology},
  volume={235},
  number={2},
  pages={R43--R61},
  year={2017},
  publisher={Bioscientifica Ltd}
}

@article{m2011diagnostic,
  title={Diagnostic and prognostic markers in differentiated thyroid cancer},
  author={M Gomez Saez, Jose},
  journal={Current Genomics},
  volume={12},
  number={8},
  pages={597--608},
  year={2011},
  publisher={Bentham Science Publishers}
}

@article{bikas2020cytochrome,
  title={Cytochrome C Oxidase Subunit 4 (COX4): A Potential Therapeutic Target for the Treatment of Medullary Thyroid Cancer},
  author={Bikas, Athanasios and Jensen, Kirk and Patel, Aneeta and Costello, John and Reynolds, Sarah M and Mendonca-Torres, Maria Cecilia and Thakur, Shilpa and Klubo-Gwiezdzinska, Joanna and Ylli, Dorina and Wartofsky, Leonard and others},
  journal={Cancers},
  volume={12},
  number={9},
  pages={2548},
  year={2020},
  publisher={Multidisciplinary Digital Publishing Institute}
}

@article{pan2001cytochrome,
  title={Cytochrome c release is upstream to activation of caspase-9, caspase-8, and caspase-3 in the enhanced apoptosis of anaplastic thyroid cancer cells induced by manumycin and paclitaxel},
  author={Pan, Jingxuan and Xu, Guangpu and Yeung, Sai-Ching Jim},
  journal={The Journal of Clinical Endocrinology \& Metabolism},
  volume={86},
  number={10},
  pages={4731--4740},
  year={2001},
  publisher={Oxford University Press}
}

@article{selinummi2009bright,
  title={Bright field microscopy as an alternative to whole cell fluorescence in automated analysis of macrophage images},
  author={Selinummi, Jyrki and Ruusuvuori, Pekka and Podolsky, Irina and Ozinsky, Adrian and Gold, Elizabeth and Yli-Harja, Olli and Aderem, Alan and Shmulevich, Ilya},
  journal={PloS one},
  volume={4},
  number={10},
  pages={e7497},
  year={2009},
  publisher={Public Library of Science San Francisco, USA}
}

@article{tabata2020bad,
  title={A bad arm existence checking problem: How to utilize asymmetric problem structure?},
  author={Tabata, Koji and Nakamura, Atsuyoshi and Honda, Junya and Komatsuzaki, Tamiki},
  journal={Machine learning},
  volume={109},
  number={2},
  pages={327--372},
  year={2020},
  publisher={Springer}
}

@article{qi2012parallel,
  title={Parallel Raman microspectroscopy using programmable multipoint illumination},
  author={Qi, Ji and Shih, Wei-Chuan},
  journal={Optics letters},
  volume={37},
  number={8},
  pages={1289--1291},
  year={2012},
  publisher={Optical Society of America}
}

@article{qi2013high,
  title={High-speed hyperspectral Raman imaging for label-free compositional microanalysis},
  author={Qi, Ji and Li, Jingting and Shih, Wei-Chuan},
  journal={Biomedical Optics Express},
  volume={4},
  number={11},
  pages={2376--2382},
  year={2013},
  publisher={Optical Society of America}
}

@article{menendez2007fatty,
  title={Fatty acid synthase and the lipogenic phenotype in cancer pathogenesis},
  author={Menendez, Javier A and Lupu, Ruth},
  journal={Nature Reviews Cancer},
  volume={7},
  number={10},
  pages={763--777},
  year={2007},
  publisher={Nature Publishing Group}
}

@article{farese2016lipid,
  title={Lipid droplets go nuclear},
  author={Farese Jr, Robert V and Walther, Tobias C},
  journal={Journal of Cell Biology},
  volume={212},
  number={1},
  pages={7--8},
  year={2016},
  publisher={The Rockefeller University Press}
}

@article{ohsaki2016pml,
  title={PML isoform II plays a critical role in nuclear lipid droplet formation},
  author={Ohsaki, Yuki and Kawai, Takeshi and Yoshikawa, Yukichika and Cheng, Jinglei and Jokitalo, Eija and Fujimoto, Toyoshi},
  journal={Journal of Cell Biology},
  volume={212},
  number={1},
  pages={29--38},
  year={2016},
  publisher={The Rockefeller University Press}
}

@article{fagin2004thyroid,
  title={How thyroid tumors start and why it matters: kinase mutants as targets for solid cancer pharmacotherapy},
  author={Fagin, JA},
  journal={Journal of Endocrinology},
  volume={183},
  number={2},
  pages={249--256},
  year={2004},
  publisher={BioScientifica}
}

@book{lennarz2013encyclopedia,
  title={Encyclopedia of biological chemistry},
  author={Lennarz, William J and Lane, M Daniel},
  year={2013},
  publisher={Academic Press}
}

@article{lopes2017regulatory,
  title={Regulatory network changes between cell lines and their tissues of origin},
  author={Lopes-Ramos, Camila M and Paulson, Joseph N and Chen, Cho-Yi and Kuijjer, Marieke L and Fagny, Maud and Platig, John and Sonawane, Abhijeet R and DeMeo, Dawn L and Quackenbush, John and Glass, Kimberly},
  journal={BMC genomics},
  volume={18},
  number={1},
  pages={1--13},
  year={2017},
  publisher={BioMed Central}
}

@article{uzunbajakava2003combined,
  title={Combined Raman and continuous-wave-excited two-photon fluorescence cell imaging},
  author={Uzunbajakava, Natallia and Otto, Cees},
  journal={Optics letters},
  volume={28},
  number={21},
  pages={2073--2075},
  year={2003},
  publisher={Optical Society of America}
}

@article{bennet2014simultaneous,
  title={Simultaneous Raman microspectroscopy and fluorescence imaging of bone mineralization in living zebrafish larvae},
  author={Bennet, M and Akiva, Anat and Faivre, D and Malkinson, Guy and Yaniv, Karina and Abdelilah-Seyfried, S and Fratzl, P and Masic, A},
  journal={Biophysical journal},
  volume={106},
  number={4},
  pages={L17--L19},
  year={2014},
  publisher={Elsevier}
}

@article{rau2019raman,
  title={Raman spectroscopy discriminates malignant follicular lymphoma from benign follicular hyperplasia and from tumour metastasis},
  author={Rau, Julietta V and Marini, Federico and Fosca, Marco and Cippitelli, Claudia and Rocchia, Massimiliano and Di Napoli, Arianna},
  journal={Talanta},
  volume={194},
  pages={763--770},
  year={2019},
  publisher={Elsevier}
}

@article{zhang2018dynamic,
  title={Dynamic sparse sampling for confocal raman microscopy},
  author={Zhang, Shijie and Song, Zhengtian and Godaliyadda, GM Dilshan P and Ye, Dong Hye and Chowdhury, Azhad U and Sengupta, Atanu and Buzzard, Gregery T and Bouman, Charles A and Simpson, Garth J},
  journal={Analytical chemistry},
  volume={90},
  number={7},
  pages={4461--4469},
  year={2018},
  publisher={ACS Publications}
}

@article{hayakawa2023lipid,
  title={Lipid droplet accumulation and adipophilin expression in follicular thyroid carcinoma},
  author={Hayakawa, Michiyo and Taylor, J Nicholas and Nakao, Ryuta and Mochizuki, Kentaro and Sawai, Yuki and Hashimoto, Kosuke and Tabata, Koji and Kumamoto, Yasuaki and Fujita, Katsumasa and Konishi, Eiichi and others},
  journal={Biochemical and biophysical research communications},
  volume={640},
  pages={192--201},
  year={2023},
  publisher={Elsevier}
}

@article{bhuiyan2023differentiability,
  title={Differentiability of cell types enhanced by detrending a non-homogeneous pattern in a line-illumination Raman microscope},
  author={Bhuiyan, Abdul Halim and Cl{\'e}ment, Jean-Emmanuel and Ferdous, Zannatul and Mochizuki, Kentaro and Tabata, Koji and Taylor, James Nicholas and Kumamoto, Yasuaki and Harada, Yoshinori and Bocklitz, Thomas and Fujita, Katsumasa and others},
  journal={Analyst},
  volume={148},
  number={15},
  pages={3574--3583},
  year={2023},
  publisher={Royal Society of Chemistry}
}

@article{delfino2019multivariate,
  title={Multivariate analysis of difference raman spectra of the irradiated nucleus and cytoplasm region of SH-SY5Y human neuroblastoma cells},
  author={Delfino, Ines and Ricciardi, Valerio and Manti, Lorenzo and Lasalvia, Maria and Lepore, Maria},
  journal={Sensors},
  volume={19},
  number={18},
  pages={3971},
  year={2019},
  publisher={MDPI}
}

@article{ivers2014dynamic,
  title={Dynamic and influential interaction of cancer cells with normal epithelial cells in 3D culture},
  author={Ivers, Laura P and Cummings, Brendan and Owolabi, Funke and Welzel, Katarzyna and Klinger, Rut and Saitoh, Sayaka and O’Connor, Darran and Fujita, Yasuyuki and Scholz, Dimitri and Itasaki, Nobue},
  journal={Cancer cell international},
  volume={14},
  number={1},
  pages={1--16},
  year={2014},
  publisher={BioMed Central}
}

@book{mody2018diagnostic,
  title={Diagnostic pathology: cytopathology},
  author={Mody, Dina R and Thrall, Michael J and Krishnamurthy, Savitri},
  year={2018},
  publisher={Elsevier}
}

@article{carvalho2017raman,
  title={Raman spectroscopic analysis of oral cells in the high wavenumber region},
  author={Carvalho, Luis Felipe CS and Bonnier, Franck and Tellez, Claudio and Dos Santos, Laurita and O'Callaghan, Kate and O'Sullivan, Jeff and Soares, Luis Eduardo S and Flint, Stephen and Martin, Airton A and Lyng, Fiona M and others},
  journal={Experimental and molecular pathology},
  volume={103},
  number={3},
  pages={255--262},
  year={2017},
  publisher={Elsevier}
}

@book{foley1996computer,
  title={Computer graphics: principles and practice},
  author={Foley, James D},
  volume={12110},
  year={1996},
  publisher={Addison-Wesley Professional}
}

@article{liu2017raman,
  title={Raman microspectroscopy of nucleus and cytoplasm for human colon cancer diagnosis},
  author={Liu, Wenjing and Wang, Hongbo and Du, Jingjing and Jing, Chuanyong},
  journal={Biosensors and Bioelectronics},
  volume={97},
  pages={70--74},
  year={2017},
  publisher={Elsevier}
}

@article{chicco2020advantages,
  title={The advantages of the Matthews correlation coefficient (MCC) over F1 score and accuracy in binary classification evaluation},
  author={Chicco, Davide and Jurman, Giuseppe},
  journal={BMC genomics},
  volume={21},
  number={1},
  pages={1--13},
  year={2020},
  publisher={BioMed Central}
}

@article{wang2023differentiating,
  title={Differentiating follicular thyroid carcinoma and thyroid adenoma by using near-infrared surface-enhanced Raman spectroscopy},
  author={Wang, Si-si and Xie, Chao and Ye, Dao-xiong and Jin, Biao},
  journal={Indian Journal of Surgery},
  pages={1--9},
  year={2023},
  publisher={Springer}
}

@article{tabata2024fly,
  title={On-the-fly Raman microscopy guaranteeing the accuracy of discrimination},
  author={Tabata, Koji and Kawagoe, Hiroyuki and Taylor, J Nicholas and Mochizuki, Kentaro and Kubo, Toshiki and Clement, Jean-Emmanuel and Kumamoto, Yasuaki and Harada, Yoshinori and Nakamura, Atsuyoshi and Fujita, Katsumasa and others},
  journal={Proceedings of the National Academy of Sciences},
  volume={121},
  number={12},
  pages={e2304866121},
  year={2024},
  publisher={National Acad Sciences}
}

@article{zink2004nuclear,
  title={Nuclear structure in cancer cells},
  author={Zink, Daniele and Fischer, Andrew H and Nickerson, Jeffrey A},
  journal={Nature reviews cancer},
  volume={4},
  number={9},
  pages={677--687},
  year={2004},
  publisher={Nature Publishing Group UK London}
}

@article{fischer2020nuclear,
  title={Nuclear morphology and the biology of cancer cells},
  author={Fischer, Edgar G},
  journal={Acta cytologica},
  volume={64},
  number={6},
  pages={511--519},
  year={2020},
  publisher={S. Karger AG}
}

@incollection{singh2022nuclear,
  title={Nuclear Morphological abnormalities in Cancer: a search for unifying mechanisms},
  author={Singh, Ishita and Lele, Tanmay P},
  booktitle={Nuclear, chromosomal, and genomic architecture in biology and medicine},
  pages={443--467},
  year={2022},
  publisher={Springer}
}

@article{duraipandian2018raman,
  title={Raman spectroscopic detection of high-grade cervical cytology: Using morphologically normal appearing cells},
  author={Duraipandian, Shiyamala and Traynor, Damien and Kearney, Padraig and Martin, Cara and O’Leary, John J and Lyng, Fiona M},
  journal={Scientific reports},
  volume={8},
  number={1},
  pages={15048},
  year={2018},
  publisher={Nature Publishing Group UK London}
}

@article{traynor2021raman,
  title={Raman spectral cytopathology for cancer diagnostic applications},
  author={Traynor, Damien and Behl, Isha and O’dea, Declan and Bonnier, Franck and Nicholson, Siobhan and O’connell, Finbar and Maguire, Aoife and Flint, Stephen and Galvin, Sheila and Healy, Claire M and others},
  journal={Nature Protocols},
  volume={16},
  number={7},
  pages={3716--3735},
  year={2021},
  publisher={Nature Publishing Group UK London}
}

@article{o2021automated,
  title={Automated raman micro-spectroscopy of epithelial cell nuclei for high-throughput classification},
  author={O’Dwyer, Kevin and Domijan, Katarina and Dignam, Adam and Butler, Marion and Hennelly, Bryan M},
  journal={Cancers},
  volume={13},
  number={19},
  pages={4767},
  year={2021},
  publisher={MDPI}
}

@article{harris2009raman,
  title={Raman spectroscopy and advanced mathematical modelling in the discrimination of human thyroid cell lines},
  author={Harris, Andrew T and Garg, Manjree and Yang, Xuebin B and Fisher, Sheila E and Kirkham, Jennifer and Smith, D Alastair and Martin-Hirsch, Dominic P and High, Alec S},
  journal={Head \& neck oncology},
  volume={1},
  pages={1--6},
  year={2009},
  publisher={Springer}
}

@article{sbroscia2020thyroid,
  title={Thyroid cancer diagnosis by Raman spectroscopy},
  author={Sbroscia, Marco and Di Gioacchino, Michael and Ascenzi, Paolo and Crucitti, Pierfilippo and di Masi, Alessandra and Giovannoni, Isabella and Longo, Filippo and Mariotti, Davide and Naciu, Anda Mihaela and Palermo, Andrea and others},
  journal={Scientific Reports},
  volume={10},
  number={1},
  pages={13342},
  year={2020},
  publisher={Nature Publishing Group UK London}
}

@article{asa2019current,
  title={The current histologic classification of thyroid cancer},
  author={Asa, Sylvia L},
  journal={Endocrinology and Metabolism Clinics},
  volume={48},
  number={1},
  pages={1--22},
  year={2019},
  publisher={Elsevier}
}

@article{mochizuki2023high,
  title={High-throughput line-illumination Raman microscopy with multislit detection},
  author={Mochizuki, Kentaro and Kumamoto, Yasuaki and Maeda, Shunsuke and Tanuma, Masato and Kasai, Atsushi and Takemura, Masashi and Harada, Yoshinori and Hashimoto, Hitoshi and Tanaka, Hideo and Smith, Nicholas Isaac and others},
  journal={Biomedical Optics Express},
  volume={14},
  number={3},
  pages={1015--1026},
  year={2023},
  publisher={Optica Publishing Group}
}

@article{wang2022diagnosis,
  title={Diagnosis accuracy of Raman spectroscopy in the diagnosis of breast cancer: a meta-analysis},
  author={Wang, Mei-Huan and Liu, Xiao and Wang, Qian and Zhang, Hua-Wei},
  journal={Analytical and Bioanalytical Chemistry},
  volume={414},
  number={27},
  pages={7911--7922},
  year={2022},
  publisher={Springer}
}

@article{jabarkheel2022rapid,
  title={Rapid intraoperative diagnosis of pediatric brain tumors using Raman spectroscopy: A machine learning approach},
  author={Jabarkheel, Rashad and Ho, Chi-Sing and Rodrigues, Adrian J and Jin, Michael C and Parker, Jonathon J and Mensah-Brown, Kobina and Yecies, Derek and Grant, Gerald A},
  journal={Neuro-Oncology Advances},
  volume={4},
  number={1},
  pages={vdac118},
  year={2022},
  publisher={Oxford University Press US}
}

@article{ohsawa2012mitochondrial,
  title={Mitochondrial defect drives non-autonomous tumour progression through Hippo signalling in Drosophila},
  author={Ohsawa, Shizue and Sato, Yoshitaka and Enomoto, Masato and Nakamura, Mai and Betsumiya, Aya and Igaki, Tatsushi},
  journal={Nature},
  volume={490},
  number={7421},
  pages={547--551},
  year={2012},
  publisher={Nature Publishing Group UK London}
}

\newpage

\appendix
\setcounter{figure}{0}
\setcounter{table}{0}
\renewcommand{\thefigure}{S\arabic{figure}}
\renewcommand{\thetable}{S\arabic{table}}

{\centering \Huge \textbf{Supplementary Materials}}
\vspace{0.5cm}

\section{Supplementary Text}

\subsection{Detailed Data Preprocessing}
\label{sup_method}
\noindent Unless specified, all the data used in our analysis were processed with the following step-by-step protocol. Note that the code to process Raman hyperspectral images with the steps below is made publicly available at \url{https://github.com/AI-SysBio/Raman-Imaging-Processing}.

\begin{enumerate}
     
    \item \textbf{Cosmic rays:}\\ The spectrum measured at some pixel are altered by cosmic rays, they can be detected when the intensity at a specific wavenumber $\omega$ is at least 8 times the standard deviation $\sigma$ higher than the average intensity $\mu$ at that wavenumber: $I(\omega) \geq 8 \sigma(\omega) + \mu(\omega)$. Pixel detected with cosmic ray contamination were replaced with the mean of the intensities in the cube surrounding the outlying pixel.

    \item \textbf{Correction for irradiance profile:}\\ 
    Because the line-shaped excitation loses intensity moving away from its center, there exist a spatial variation in irradiance of the sample along the excitation axis during the measurement (Fig. 1C in~\cite{bhuiyan2023differentiability}). In our setup, the irradiance profile typically involves a 1-2\% difference in the illumination intensity between the edge and the center of the image. As we later identified background and cell regions based on intensity clustering (step 3), we needed to correct for this irradiation. With the assumption that the intensity along the excitation line follows a Gaussian profile, we estimated the irradiance profile with a Gaussian fit. Let
    $\bar{I}(x, y, \omega)$ be the Raman intensity at position $x$ on the illumination axis, $y$ the scanning axis and wavenumber $\omega$. First, we summed the irradiation over the scanning axis.

    \begin{equation}
        I(x,\omega) = \mathlarger{\sum}_y \ \bar{I}(x, y, \omega).
    \end{equation}

    Then, for each wavenumber $\omega$, we fitted a Gaussian function $I_\text{irr}(x,\omega)$ by minimizing the square error with $I(x,\omega)$.

    \begin{equation}
        I_\text{irr}(x,\omega) = a_\omega \cdot \exp\left(-\frac{(x -x_{\omega})^2}{2\sigma_\omega^2}\right) , \ \ \text{minimizing} \ \ \mathlarger{\sum}_x \Bigl[ I_\text{irr}(x,\omega) - I(x,\omega) \Bigr] ^2.
    \end{equation}

    After repetition at all wavenumber frames, the  Gaussian fits were normalized by scaling their maximum to one and averaged over all wavenumbers, yielding an estimated irradiance profile for the whole image. Wavenumbers at which the Gaussian fit resulted in a standard deviation larger than two times the image size on the illumination axis were not considered. 

    \begin{equation}
        I'_\text{irr}(x) = \mathlarger{\mathlarger{\sum}}_\omega \frac{\ I_\text{irr}(x,\omega)}{a_\omega}.
    \end{equation}
    
    Finally, we obtained the corrected Raman intensities by dividing the initial intensities by this profile.

    \begin{equation}
        I_\text{adjusted}(x,y,\omega) = \frac{I(x,y,\omega)}{I'_\text{irr}(x)}
    \end{equation}

    %
    %
    %
    %
    %

    \item \textbf{Identification of background regions and cell regions:}\\
    
    
    To produce the \textit{background mask}, the image was first partitioned into $k = 2$ groups by $k$-mean clustering on the average intensity over the high wavenumber region (2800-$\SI{3040}{cm^{-1}}$). Let $I_{i,j}$ denote the average Raman intensity at pixel $(i,j)$ over the high wavenumber region, $n_k$ the total number of pixels belonging to the cluster $k$, and $c_k$  its center:

    \begin{equation}
       c_k=\frac{1}{n_k}\mathlarger{\mathlarger{\sum}}_{(i,j)\in \mathrm{cluster~} k} I_{i,j}.
    \end{equation}
    
    The clustering was repeated by recursively increasing the number of clusters $k$, until the distance of each pixel to its corresponding cluster center was on average smaller than the distance produced by Poisson error. 
    
    \begin{equation}
       \mathlarger{\mathlarger{\sum}}_{(i,j)\in \mathrm{cluster~} k} \sqrt{\left(I_{i,j} - c_k\right)^2} \leq \mathlarger{\mathlarger{\sum}}_{(i,j)\in \mathrm{cluster~} k} \sqrt{2 I_{i,j}}.
    \label{condition}
    \end{equation}
    
     Once the condition \ref{condition} was verified for all clusters $k$, the spectra belonging to the least intense cluster were retained as pixels containing background regions.\\

     Then, the \textit{cell mask} was defined by retaining spectra belonging to the $m$ most intense clusters. $m=\lfloor k/2 \rfloor$ was chosen in our analysis (See Supplementary Section~\ref{sup:mclusters} with $k=10$). The connected regions of pixel contained in cells were identified with a flood-fill algorithm~\cite{nosal2008flood}. Isolated regions containing less than 200 pixels ($\sim  200 \mu \mathrm{m}^{2}$) were too small to be cells and were discarded. In some Raman images, some strong overlap existed between the cells because of high cell density. In these cases, cells were segmented manually. Fig.~\ref{fig:heterogenous}(c) and Figs.~\ref{fig:cancer_index}(b), \ref{fig:ground_nuc_SI}(b) illustrate the resultant background segmentation in mono-cultured, and co-cultured systems, and Fig.~\ref{fig:co-culture1} second columns show the segmented manually. Fig.~\ref{fig:heterogenous}(c) and Figs.~\ref{fig:cancer_index}(b), \ref{fig:ground_nuc_SI}(b) illustrate the resultant background, and cell segmentation for all the co-cultured images analyzed in this paper.

    \item \textbf{Superpixel:}\\
    For the purposes of further increasing the signal-to-noise ratio and reducing the computational cost, adjacent pixels were averaged over spatially-local regions within the images, producing spectra representing groups of pixels, or superpixels. Rather than using a simple binning scheme on a square grid of predetermined size and location, we instead chose the Simple Linear Iterative Clustering (SLIC)~\cite{achanta2012slic} pixel clustering method, that is based on spatial proximity and color similarity to better preserve the spatial characteristics of the cells in the Raman images. The number of superpixels per image was chosen such that the superpixels contain an average of 10 individual pixels, which provides superpixels having an average area of $\sim  1 \mu \mathrm{m}^{2}$. The \texttt{Python} implementation from \textit{scikit-image} library was used.
    
    \item \textbf{Background subtraction:}\\ For each image, the background spectra was computed by averaging the spectra of all pixels where \textit{background mask} is 1. Then, the mean background spectra was subtracted to the spectrum of all pixels in the image.
%
    Since we only have interest in the Raman analysis of cell regions. All the pixels where \textit{cell mask} is 0 were discarded.
%
    All spectra were interpolated to a common wavenumber axis and kept only in the (700-$\SI{3000}{cm^{-1}}$) wavenumber range (because the wavenumber axis may vary slightly from image to image).
    
    \item \textbf{Denoising:}\\ The intensity of each wave number in the Raman spectrum is following a Poisson distribution. Denoising was performed by keeping the first 7 components of the measurement matrix after its Singular Value Decomposition (SVD).
    
    \item \textbf{Fluorescence background:}\\ All the spectra contain a relatively large background due to glass and water Raman auto-fluorescence, the baseline removal was done with a $7^{\text{th}}$ order recursive polynomial fitting (Polyfit)\cite{lieber2003automated}.
    
    \item \textbf{Normalization:}\\ Each spectrum was divided by the sum of its intensity over the (700-$\SI{1800}{cm^{-1}}$) and (2800-$\SI{3000}{cm^{-1}}$) wavenumber range (the silent region was cropped).

\end{enumerate}

\newpage

\subsection{Sensitivity of single cell classification with respect to the number of clusters $m$ defining cell regions}

\label{sup:mclusters}

\noindent 

\noindent In Method~\ref{method:cell_identification}, we describe how cellular regions are defined by clustering pixels based on Raman intensity within the high wavenumber spectrum (2800-$\SI{3040}{cm^{-1}}$) into $n$ groups, retaining the $m$ most intense clusters that maximizes the FTC/NT cell classification performance as the cellular region. We swiped through different values of $m$ to evaluate how it affects classification performance of FTC. For each value of $m$, our evaluation process was as follows:

\begin{itemize}
    \item Cellular regions were determined by retaining the $m$ most intense clusters from the high wavenumber region after applying $k$-means clustering with $n$ clusters.
    \item Nucleus regions were identified based on the average intensity within the CH$_2$ stretching wavenumber range (2930-$\SI{3000}{cm^{-1}}$), as detailed in Method~\ref{method:nucleus_identification}.
    \item Individual nuclei were inferred using a flood-fill algorithm~\cite{foley1996computer}.
    \item Individual cells were identified by linking each cellular pixel to its nearest nucleus.
    \item Cells were classified as either FTC or NT using a 10-nearest neighbor (10-NN) algorithm with the first five principal components.
\end{itemize}

\begin{figure}[h!t]
    \centering
    \includegraphics[width=0.5\columnwidth]{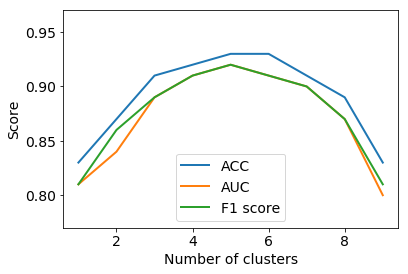}
    \caption{Accuracy (ACC), AUC and F1score obtained for the FTC/NT classification as a function of the number of clusters used for cell region identification after $k$-means clustering with 10 clusters.}
    \label{fig:cellclust}
\end{figure}

Figure~\ref{fig:cellclust} presents an example with $n = 10$ clusters, where the AUC, accuracy, and F1 scores peak at $m = 5$ clusters. After sweeping through different number of group from $n = 2$ to $20$, we found that $m = \lfloor n/2 \rfloor$ consistently maximized the AUC, emerging as the optimal choice in $90\%$ of the cases.

\subsection{Dependence of $k$ and $n$PCs for kNN classification}

\label{sup:kPC}

\begin{figure}[h!t]
    \centering
    \includegraphics[width=1\columnwidth]{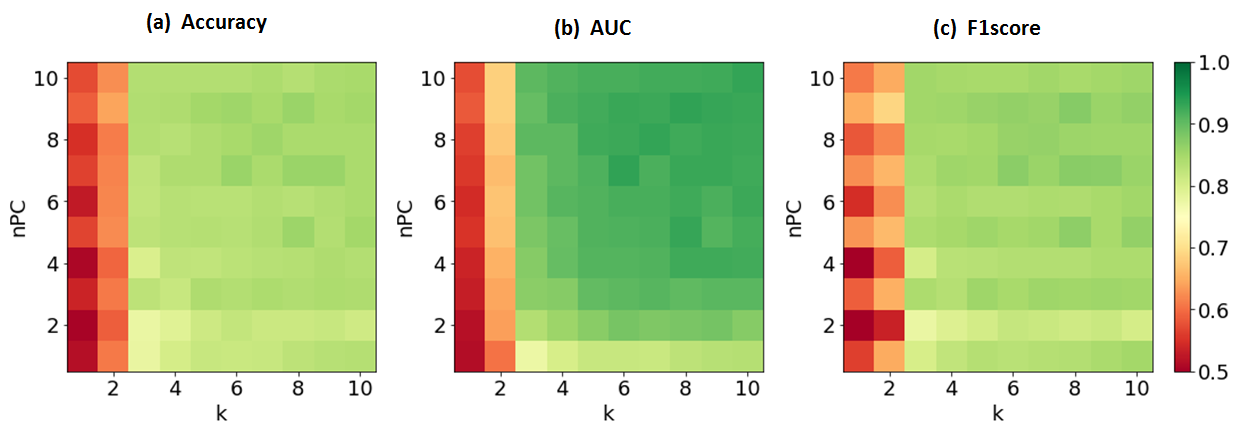}
    \caption{Accuracy, AUC and F1score as a function of number of neighbors and number of PCs. Computed with a nearest neighbor classifier 5-fold cross validation averaged over 1000 independent shuffling on average cell spectra.}
    \label{fig:k_nPC}
\end{figure}

\noindent Figure \ref{fig:k_nPC} shows that the classification performance is not significantly affected for kNN classification unless $k<3$ or $nPC<3$.

\newpage

\subsection{Statistical analysis of SV40 immunofluorescence imaging}

\label{SV40}

\noindent Two cell cultures were prepared, one with Nthy-ori 3-1 cell line, the other with FTC-133 cell line. After preparation of the samples as described in Method~\ref{sample_prep}, immunofluorescence measurements of the SV40 protein were performed in order to assign the ground truth, i.e., which cells are of FTC-133 and Nthy-ori 3-1 in the co-culture system, including the error rate of the assignment. By processing the obtained images with ImageJ, we obtained the fluorescence intensity from 872 Nthy-ori 3-1 nuclei and 566 FTC-133 nuclei. 
Setting an intensity threshold at 1.9 to identify FTC-133 and Nthy-ori 3-1 cells, we found that just two Nthy-ori 3-1 cells fell below this threshold, resulting in two false positives. Consequently, this resulted in an error rate of 2 out of 1438, which is approximately \SI{0.14}{\%}.

\begin{figure}[h!t]
    \centering
    \includegraphics[width=1\columnwidth]{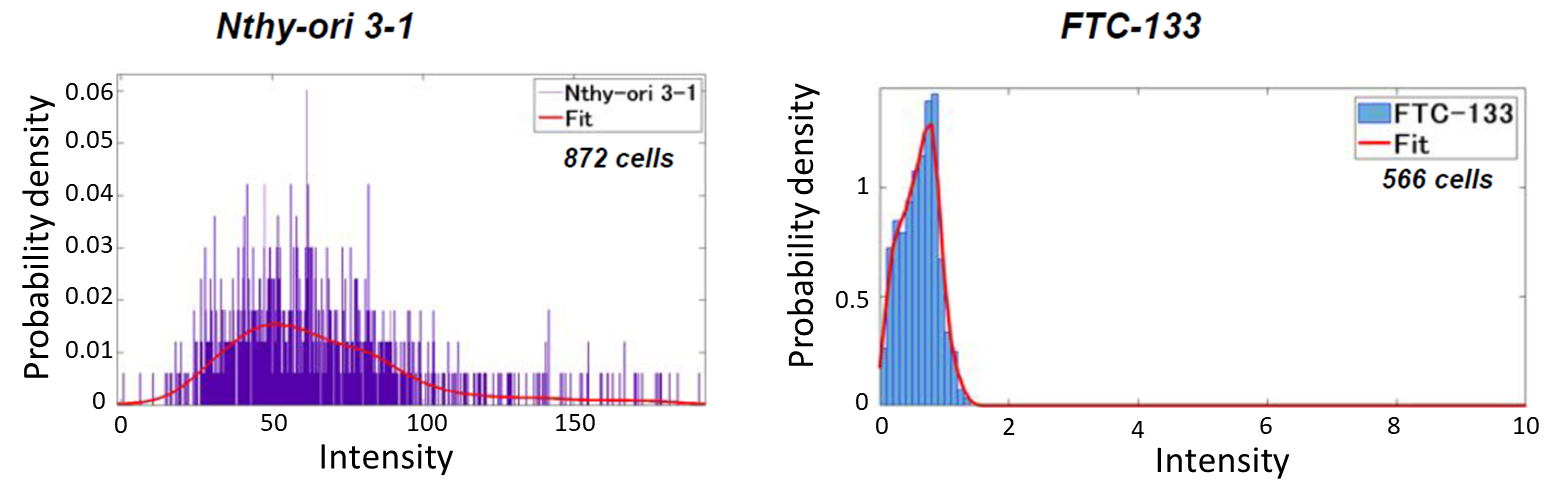}
    \caption{Nucleus fluorescent intensity distribution after SV40 staining for (a) Nthy-ori 3-1 and (b) FTC-133 cell lines. The red curve is the probability density function obtained after fitting by a Gaussian kernel.}
    \label{fig:SV40}
\end{figure}



\subsection{Co-cultured system analysis}
\noindent Table S1 provides the accuracy (Acc), false positive (FP), and false negative (FN) obtained for each images via position matching to fluorescence images. The column U (undetermined) corresponds to cells than cannot be decided, due to a strong overlap between FTC and NT cells, or because it was washed out before the fluorescence measurement. For example, for the co-cultured system named `190607\_FTC\_Nthy\_no1,' total number of single cells assigned is 52, composed of 30 FTC-133 and 20 Nthy-ori 3-1 as well as two cells for which malignancy could not be assigned. Among them 2 out of 20 Nthy-ori 3-1 cells were falsely identified as being FTC-133, and 5 out of 30 FTC-133 were falsely identified as being Nthy-ori 3-1, with 2 undetermined cells. All the corresponding figures used for the classification are provided on Figure~\ref{fig:co-culture1}.


\def\arraystretch{1.2}
\begin{table*}[ht]
\centering
\caption{Number of False Positive (FP), False Negative (FN), and undetermined (U) cells of each colculture system, the computed accuracy (Acc) is also provided. $N_\text{NT}$ and $N_\text{FTC}$ refers to the ground truth number of NT and FTC cells in the image, respectively.}
\begin{tabular}{|C{5cm}||C{2cm}|C{2cm}|C{2cm}|C{2cm}|C{2cm}|}  
    \hline
    Image & \# cells & FP/$N_\text{NT}$ & FN/$N_\text{FTC}$ & U & Acc \\
    \hline
    190607\_FTC\_Nthy\_no1 & 52 & 2/20 & 5/30 & 2 & 86\%\\ 
    \hline
    190607\_FTC\_Nthy\_no2 & 68 & 4/60 & 2/6 & 2 & 91\%\\ 
    \hline
    190607\_FTC\_Nthy\_no3 & 60 & 4/30 & 4/30 & 0 & 87\%\\
    \hline
    190614\_FTC\_Nthy\_no1 & 46 & 6/31 & 0/15 & 0 & 85\%\\ 
    \hline
    190614\_RO82\_Nthy\_no2 & 45 & 18/25 & 3/15 & 5 & 48\%\\
    \hline
    190620\_RO82\_Nthy\_no1 &  60 & 3/21 & 4/37 & 2 & 88\%\\
    \hline
    190620\_RO82\_Nthy\_no2 &  56 & 5/23 & 3/29 & 4 & 85\%\\
    \hline
    190627\_FTC\_Nthy\_no1 &  46 & 6/26 & 1/19 & 1 & 84\%\\
    \hline
    190627\_FTC\_Nthy\_no2 &  47 & 4/25 & 1/20 & 2 & 89\%\\
    \hline
    Total &  480 & 51/261 & 24/201 & 18 & \SI{83.8}{\%}\\
    \hline

\end{tabular}
\label{table:co-culture}
\end{table*}

\def\arraystretch{1}

\newpage

\section{Supplementary Tables}

\def\arraystretch{1.6}
\begin{table*}[h!t]
\centering
\caption{Cellular classification accuracy, AUC and F1 score with a 10 Nearest Neighbor classifier on the first 5 principal components of the average cellular spectra with 5-fold cross validation averaged over 1000 sampling. Evaluation was performed by excluding (ex.) the cells measured with the same device, at the same date and from the same image, respectively. Classification performance without exclusion is also provided.}
\begin{tabular}{|c|C{4cm}||C{2.5cm}|C{2.5cm}|C{2.5cm}|C{2.5cm}|}  
    \cline{3-6}
    \multicolumn{2}{c||}{} & Ex. by device & Ex. by date & Ex. by image & No exclusion\\  
    \hline
    \multirow{3}{*}{\STAB{\rotatebox[origin=c]{90}{Without EBC}}}
    & Acc & 68 $\pm$ 0.9\% & 77 $\pm$ 0.9\% & 80 $\pm$ 1.1\% & 83 $\pm$ 1.3\%\\
    \cline{2-6}
    & AUC &  0.77 $\pm$ 0.008 & 0.86 $\pm$ 0.009 & 0.90 $\pm$ 0.009 & 0.91 $\pm$ 0.011\\
    \cline{2-6}
    & F1score  &  0.73 $\pm$ 0.008 & 0.80 $\pm$ 0.009 & 0.81 $\pm$ 0.009 & 0.85 $\pm$ 0.010\\
    \hline
    \hline
    \multirow{3}{*}{\STAB{\rotatebox[origin=c]{90}{With EBC}}}
    & Acc & 81 $\pm$ 1.0\% & 81 $\pm$ 1.0\% & 81 $\pm$ 1.1\% & 83 $\pm$ 1.2\%\\ 
    \cline{2-6}
    & AUC & 0.89 $\pm$ 0.010 & 0.89 $\pm$ 0.010 & 0.90 $\pm$ 0.010 & 0.91 $\pm$ 0.011\\
    \cline{2-6}
    & F1score & 0.82 $\pm$ 0.010 & 0.82 $\pm$ 0.010 & 0.82 $\pm$ 0.010 & 0.84 $\pm$ 0.011\\
    \hline
\end{tabular}
\label{table:classif_prot}
\end{table*}

\def\arraystretch{1.}

\def\arraystretch{1.2}
\begin{table*}[ht]
\centering
\caption{ Assignment of the Major Peaks in the Raman Spectra of FTC and NT cell lines.\\ $\beta$ = bending; $\nu$ = stretching (s = symmetric; as = asymmetric)}
\label{table:wavenumbers}
\resizebox{\textwidth}{!}{
\begin{tabular}{|C{3cm}|C{4.2cm}|C{4.4cm}|C{4.8cm}|}  

    \hline
    Wavenumber peak ($\pm \SI{5}{cm^{-1}}$) & Amino acids~\cite{rygula2013raman, rau2016raman} \mbox{\small (building blocks of proteins)}  & Lipids\cite{czamara2015raman} \mbox{\small(Fatty acids, Triacylglycerols)} &  Others  \mbox{\small(DNA~\cite{de2007reference}, Cytochromes~\cite{okada2012label})}\\
    
    \hline
    \hline
    723 &  & & Adenine (DNA) \\
    \hline
    750 &  Tryptophan & & Cytochromes\\
    \hline
    786 & & & Cytosine (DNA) \\ 
    \hline
    828 & Tyrosine & &\\
    \hline
    855 & Tyrosine, Proline & &\\
    \hline
    938 & Proline & & \\
    \hline
    1004 & Phenylalanine  & &\\
    \hline
    1032 & Phenylalanine & &\\
    \hline
    1068 and 1089 & $\nu(\text{C}-\text{C})$ & $\nu(\text{C}-\text{C})$& \\
    \hline
    1127 & $\nu(\text{C}-\text{N})$ & & Cytochromes\\
    \hline
    1155 & $\nu(\text{C}-\text{C})$ and $\nu(\text{C}-\text{N})$& $\nu(\text{C}-\text{C})$& \\
    \hline
    1175 & Tyrosine & &\\
    \hline
    1211 & Amide III & & \\
    \hline
    1265 & Amide III & & \\
    \hline
    1310 & Tryptophan & & Cytochromes\\
    \hline
    1339 & Tryptophan, Amide III & & \\
    \hline
    1447 &  $\nu(\text{C}-\text{H})$ & $\beta(\text{CH}_2) ,\beta(\text{CH}_3)$ &\\
    \hline
    1581 & Phenylalanine, Proline & & Cytochromes\\
    \hline
    1660 & Amide I & $\nu_\text{s}(\text{C}=\text{C})$ &\\    
    \hline
    2854 & & $\nu_\text{s}(=\text{CH}_2)$  &\\
    \hline
    2890 & & $\nu_\text{as}(=\text{CH}_2)$  &\\
    \hline
    2934 & $\nu_\text{s}(=\text{CH}_3)$ & $\nu_\text{s}(=\text{CH}_3)$  & $\nu_\text{s}(=\text{CH}_3)$\\
    \hline
\end{tabular}
}
\label{table:Ramanshift}
\end{table*}

\def\arraystretch{1.}

\def\arraystretch{1.2}
\begin{table*}[ht]
\centering
\caption{Experimental setup details of device 1 and device 2 (line illumination Raman microscope). The co-culture system measurements were performed with both device and an 20x/0.75 (UPLSAPO, Olympus) objective lens.}
\begin{tabular}{|C{4cm}|C{6cm}|C{6cm}|}  
    \cline{2-3}
    \multicolumn{1}{c|}{} & Device 1 (commercial system) & Device 2 (home-built system) \\
    \hline
    Excitation light source & \SI{532}{nm} CW laser & \SI{532}{nm} CW laser (Coherent, Verdi V18) \\
    \hline
    Objective lens &  60x/1.2, water immersion (Olympus, UPLSAPO 60XW)  &  40x/1.25 (Nikon, CFI Apo 40xWI $\lambda$S) \\
    \hline
    Edge filter &  - & Semrock, LP03-532RE-25 \\
    \hline
    Spectrophotometer &  - & Bunkokeiki, MK-300 \\
    \hline
    Grating &  \SI{600}{g/mm} & \SI{600}{g/mm} \\
    \hline
    Detector &  Cooled CCD (Princeton instruments, PIXIS400B eXcelon) & Cooled CCD (Princeton instruments, PIXIS400B eXcelon) \\
    \hline
    Slit width &  $\SI{30}{\micro\metre}$ & $\SI{30}{\micro\metre}$  \\
    \hline
    Illuminated line width at the slit plane & $\SI{32}{\micro\metre}$ & $\SI{25}{\micro\metre}$  \\
    \hline
    Pixel resolution on wavenumber axis & $\sim \SI{2.3}{cm^{-1}/pixel}$ & $\sim \SI{3.2}{cm^{-1}/pixel}$ \\
    \hline
    Laser power &  $\SI{2.7}{mW / \micro\metre^2}$  & $\SI{3.3}{mW / \micro\metre^2}$ \\
    \hline
    Exposure time &  $\SI{10}{s}$ & $\SI{5}{s}$ \\
    \hline
    Image size &  $\SI{122}{\micro\metre} \times (77-107)\SI{}{\micro\metre}$ & $\SI{130}{\micro\metre} \times \SI{80}{\micro\metre}$ \\
    \hline
    Sampling &  $400 \times (250 - 350)$ pixels  & $400 \times 240$ pixels  \\
    \hline
\end{tabular}
\label{table:setup}
\end{table*}

\def\arraystretch{1.}




\clearpage
\newpage

\section{Supplementary Figures}


\begin{figure}[h!t]
    \centering
    \includegraphics[width=0.85\columnwidth]{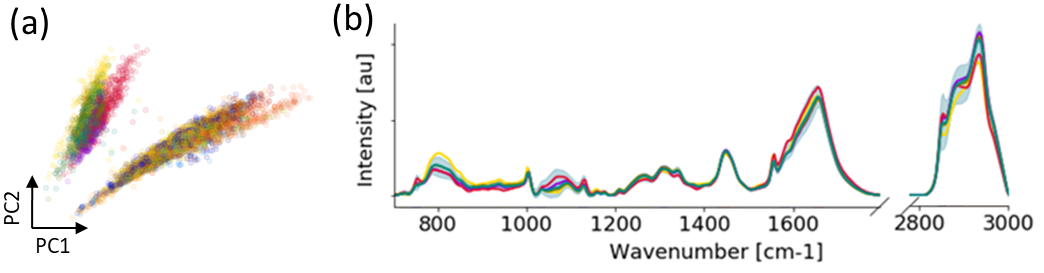}
    \caption{(a) Scatter plot of the first two principal components (capturing 88\% of the variance) of each spectrum colored by their date. (b) Raman spectra averaged over all pixels at each date of device 2. The gray area highlights the standard deviation of all spectra from device 2. }
    \label{fig:SI_dates}
\end{figure}

\begin{figure}[h!t]
    \centering
    \includegraphics[width=1\columnwidth]{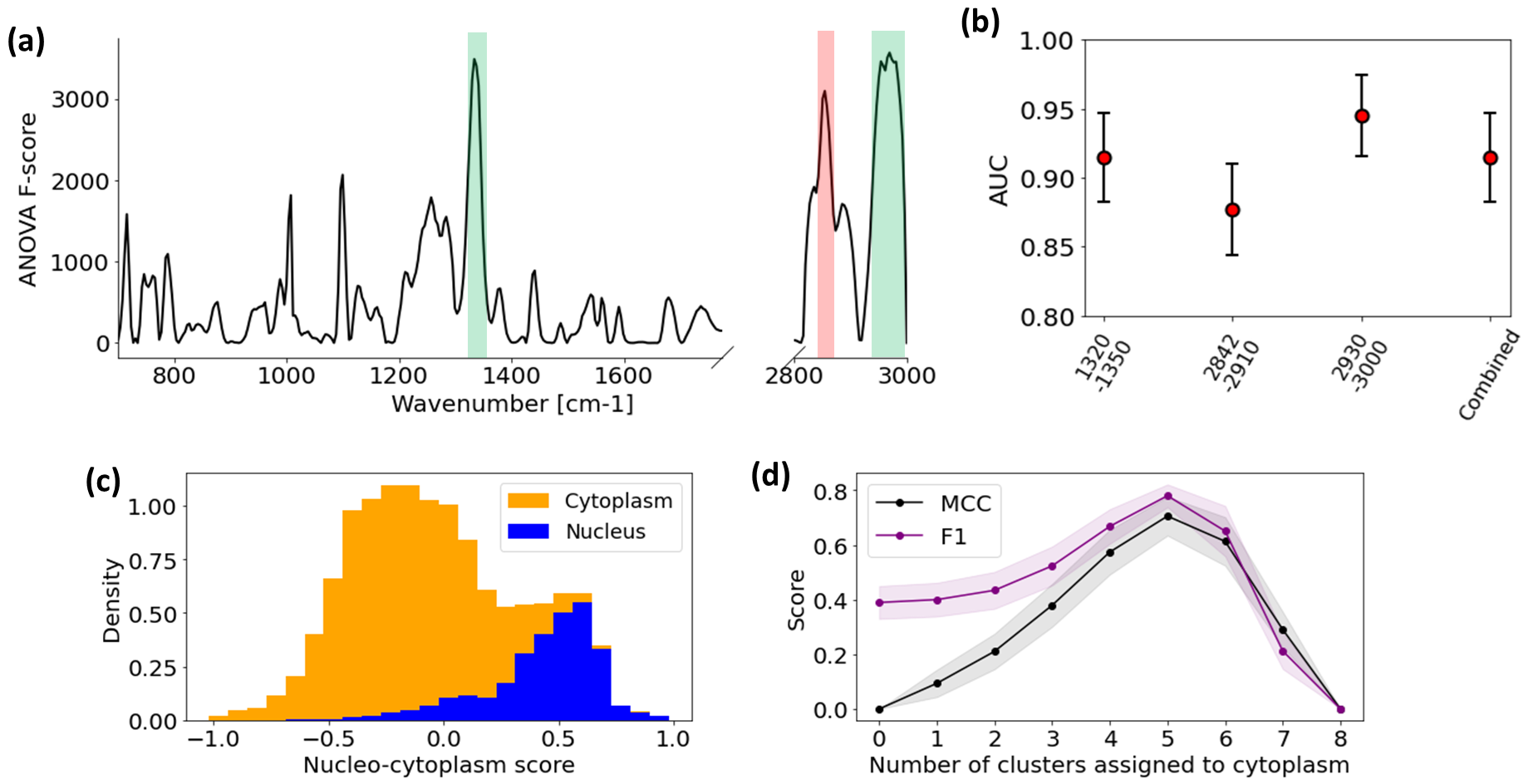}
    \caption{(a) ANOVA $F$-score of each Raman shift, wavenumber bands most relevant to the nucleus/cytoplasm differentiation are highlighted in green and red, where green indicate a positive difference and red a negative difference (as referenced in Figure~\ref{fig:nuc_identification}d). (b) Nucleus/cytoplasm classification AUC by using the mean wavenumber intensity of the three wavenumber bands with a high ANOVA score. Different combination of the bands were also tested. (c) Distribution of average Raman intensity over the $2930 - \SI{3000}{cm^{-1}}$ band of cellular regions, where contribution to the density is plotted in different colors depending if the spectra is from nucleus or cytoplasm region. (d) F1 and Matthews correlation coefficient (MCC)~\cite{chicco2020advantages} score of the identified nucleus/cytoplasm regions depending on the number of clusters associated to cytoplasm region after $k$-mean clustering with 8 clusters (note that the AUC is independent of such threshold).}
    \label{fig:ground_nuc_SI}
\end{figure}

\begin{figure}[h!t]
    \centering
    \includegraphics[width=1\columnwidth]{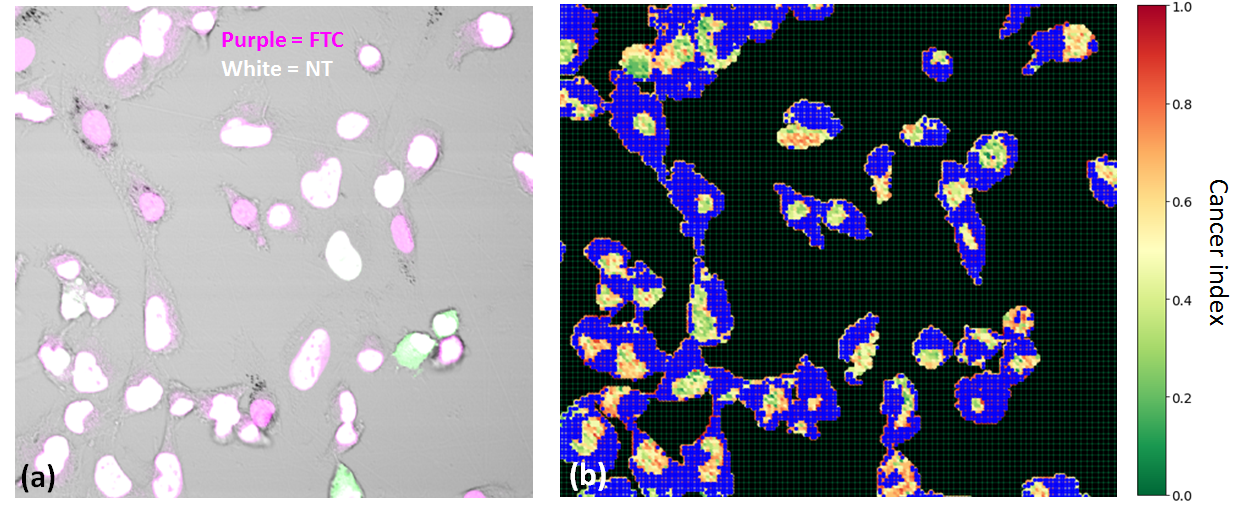}
    \caption{A representative image of the co-culture system of FTC-133 and Nthy-ori 3-1. (a) An overlaid microscopic fluorescence image of SV40 large T protein (green) and double strand DNA (magenta). Since Nthy-ori 3-1 should be positive for both, the color of nuclei in Nthy-ori 3-1 is white as a result of the overlay, in contrast, FTC cell is negative for SV40, which makes nuclei of FTC cells colored in magenta. Cells in which green color around white indicates dividing Nthy-ori 3-1.  (b) A distribution of the cancer index at each pixel trained by only nucleus Raman signals, pixels from the background and cytoplasm regions are colored in black and blue, respectively. The cancer index is close to 0.5 for all nucleus, illustrating the low performances of nucleus classification.}
    \label{fig:cancer_index_nuc}
\end{figure}




\begin{figure}[h!t]
    \centering
    \includegraphics[width=0.9\columnwidth]{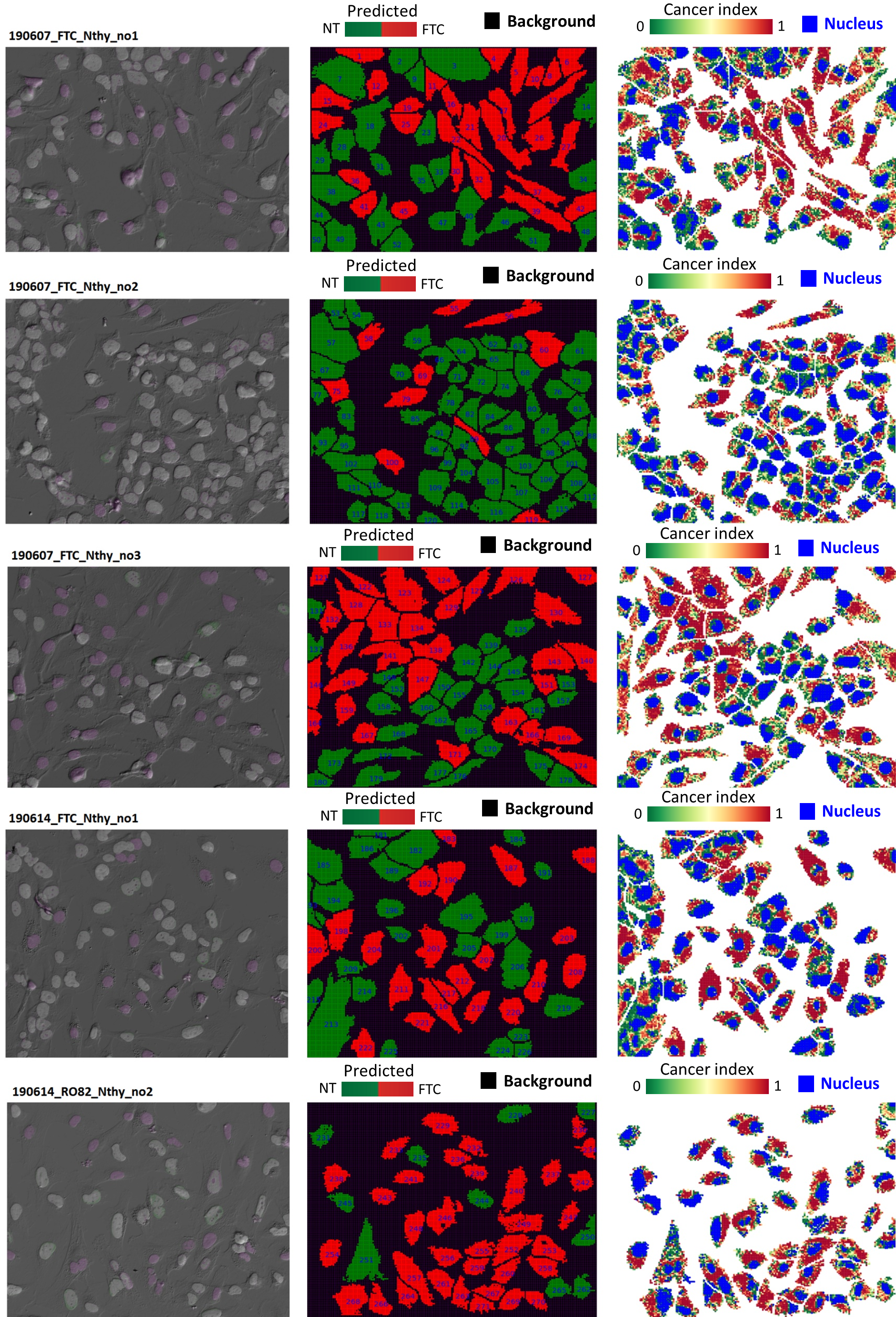}
    \caption{ (left) Immunofluorescence imaging to discriminate (ground truth) between FTC (FTC-133, RO82W-1) and NT (Nthy-ori 3-1) cells by SV40 large T antigen protein. Cells colored in purple (white) corresponds to FTC (NT).  (middle) The prediction of FTC/NT by introducing a threshold in the averaged cancer index taken over each single cell, selected to maximize the F1 score of our training data. Cells colored in red (green) are predicted to be FTC (NT). (right) the computed cancer index at each pixel for each Raman image in the co-culture system.}
    \label{fig:co-culture1}
\end{figure}

\setcounter{figure}{5}
\begin{figure}[h!t]
    \centering
    \includegraphics[width=0.9\columnwidth]{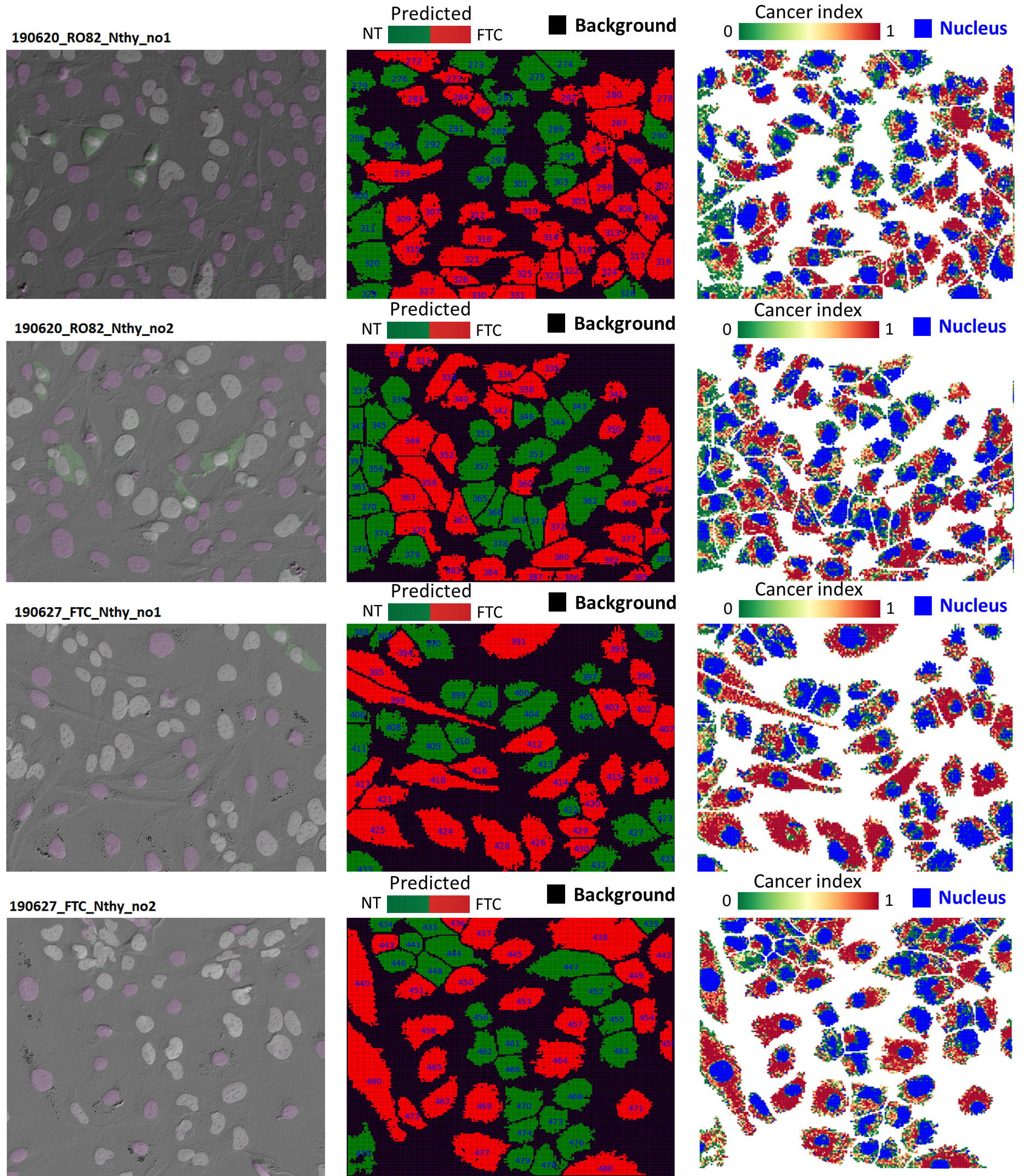}
    \caption{(Continued)}
    \label{fig:co-culture2}
\end{figure}

\end{document}